\documentstyle[preprint,aps]{revtex}

\tighten
\begin{document}

\title{
       Four-nucleon shell-model calculations in a Faddeev-like approach 
}
\medskip

\author{
        P. Navr\'atil\footnote{On leave of absence from the
   Institute of Nuclear Physics,
                   Academy of Sciences of the Czech Republic,
                   250 68 \v{R}e\v{z} near Prague,
                     Czech Republic.} and
     B. R. Barrett
        }

\medskip

\address{
                   Department of Physics,
                   University of Arizona,
                   Tucson, Arizona 85721
}

\maketitle

\bigskip

\begin{abstract}
We use equations for Faddeev amplitudes to solve the shell-model 
problem for four nucleons in the model space that includes
up to $14 \hbar\Omega$ harmonic-oscillator excitations above 
the unperturbed ground state. 
Two- and three-body effective interactions derived from the 
Reid93 and Argonne V8' nucleon-nucleon potentials are used 
in the calculations. 
Binding energies, excitations energies, point-nucleon radii
and electromagnetic and strangeness charge form factors for $^4$He
are studied.
The structure of the Faddeev-like equations is discussed and 
a formula for matrix elements of the permutation operators 
in a harmonic-oscillator basis is given.
The dependence on harmonic-oscillator excitations
allowed in the model space
and on the harmonic-oscillator frequency is investigated.
It is demonstrated that the
use of the three-body effective interactions improves the convergence
of the results.   
\end{abstract}

\bigskip
\bigskip
\bigskip

\narrowtext



\section{Introduction}
\label{sec1}

Many different methods have been used to solve the 
few-body problem in the past. One of the most viable 
approaches appears to be the Faddeev method \cite{Fad60}.
It has been successfully applied to solve the three-nucleon bound-state
problem for various nucleon-nucleon 
potentials \cite{PFGA80,PFG80,CPFG85,FPSS93,NHKG97}.
For solution of the four-nucleon problem 
one can employ Yakubovsky's generalization
of the Faddeev formalism  \cite{Ya67} as done, e.g., in Ref. \cite{GH93}.
Alternatively, other methods have also been succesfully used in the past,
such as, the Green's function Monte Carlo method \cite{GFMC}
or the correlated hyperspherical harmonics expansion method \cite{VKR95}.  

On the other hand, when studying the properties of more
complex nuclei one typically resorts to the shell model.
In that approach, the harmonic-oscillator basis is used in a truncated 
model space. Instead of the free nucleon-nucleon potential, 
one utilizes effective interactions appropriate for the truncated 
model space. 
Examples of such calculations are the
large-basis no-core shell-model calculations that have recently been 
performed \cite{ZBVHS,NB96}. 
In these calculations all nucleons are active,
which simplifies the effective interaction as no hole states
are present. The effective interaction is
determined for a system of two nucleons in a harmonic-oscillator well
interacting by the nucleon-nucleon potential and is subsequently
used in the many-particle calculations.

In a recent paper we combined the shell-model approach 
to the three-nucleon problem with the
Faddeev formalism \cite{NB98}. That allowed us to extend the shell-model
calculations to a model space of excitations of $32\hbar\Omega$
above the unperturbed ground state and to study
the convergence with respect to the size of the model space.  
In the present paper we generalize these earlier calculations 
to the four-nucleon problem. We introduce equations for Faddeev
amplitudes that are fully antisymmetrized for three nucleons.
As the center-of-mass term is removed, we are able to work 
in a model space up to an excitation of $14\hbar\Omega$ 
above the unperturbed ground state. For comparison
the largest shell-model calculations so far for $^4$He 
are those performed
by R. Ceuleneer et al., in which a $10\hbar\Omega$ model 
space was utilized \cite{CVS88}.
The main motivation for the present work is to test
the shell-model approach and the effective interactions 
derived from realistic nucleon-nucleon (NN) potentials
that are used in the conventional shell-model applications 
for more complex systems. 
As the equations that we employ can be conveniently 
used with three-body interactions or three-body effective interactions,
we investigate, in addition to two-body effective interactions, also 
three-body effective interactions in 
the present formalism. Such effective interactions are not typically used 
in the traditional applications. We show that the inclusion
of three-body effective interactions improves the overall convergence
of the results. At the same time our work serves as an alternative
method to solving the four-nucleon problem. 
We can study the convergence properties of the results with the
increasing size of the model space.
If convergence is 
achieved, our results will approach the exact solutions obtained by other
methods. In our formalism we seek simultaneously solutions 
for both the ground-state
and the excited states. In the past, the variational Monte Carlo
method was used to investigate the excited states of $^4$He 
using realistic NN potentials \cite{CPW84}.
In most four-nucleon calculations
with realistic NN potentials, however, only the ground-state 
properties were evaluated \cite{GH93,VKR95}. On the other hand,
earlier studies that investigated the excited-state properties 
usually did not employ realistic NN potentials 
\cite{CVS88,CH97}. Recently, the four-nucleon resonant and
scattering states were investigated using realistic NN potentials 
in the framework of the resonating group method \cite{HH97},
the correlated-hyperspherical-harmonics method \cite{VRK98} 
as well as in the solution of the Faddeev-Yakubovsky 
equations in configuration space \cite{CC98}.

The present calculation is simplified by using a compact formula
for the matrix elements of the permutation operators
in the harmonic-oscillator (HO) basis.
Also, because of the way we do the model-space truncation, we keep
equivalence of the Faddeev-like and Schr\"odinger equations throughout 
the calculation. In addition to calculation of ground-state
and excited state energies and point-nucleon rms radii, we also
evaluate electromagnetic (EM) and strangeness form factors
in the impulse approximation.

In section \ref{sec2} we first discuss 
the Faddeev equations for the shell-model problem of 
three nucleons. Then  
a generalization to the four-nucleon system is introduced. In
section \ref{sec3} we present the energy, radii and form factor
results for $^4$He. 
Conclusions are given in section \ref{sec4}.

\section{Shell model and Faddeev-like formalism}
\label{sec2}

In shell-model studies the one- plus two-body Hamiltonian
for the A-nucleon system, i.e.,
\begin{equation}\label{ham}
H=\sum_{i=1}^A \frac{\vec{p}_i^2}{2m}+\sum_{i<j}^A 
V_{\rm N}(\vec{r}_i-\vec{r}_j) \; ,
\end{equation}
where $m$ is the nucleon mass and $V_{\rm N}(\vec{r}_i-\vec{r}_j)$, 
the NN interaction,
is usually modified by adding the center-of-mass HO potential
$\frac{1}{2}Am\Omega^2 \vec{R}^2$, 
$\vec{R}=\frac{1}{A}\sum_{i=1}^{A}\vec{r}_i$.
This potential does not influence intrinsic properties of the 
many-body system. It provides, however, a mean field felt by each nucleon
and allows us to work with a convenient HO basis.
The modified Hamiltonian, depending on the HO 
frequency $\Omega$, can be cast into the form
\begin{equation}\label{hamomega}
H^\Omega=\sum_{i=1}^A \left[ \frac{\vec{p}_i^2}{2m}
+\frac{1}{2}m\Omega^2 \vec{r}^2_i
\right] + \sum_{i<j}^A \left[ V_{\rm N}(\vec{r}_i-\vec{r}_j)
-\frac{m\Omega^2}{2A}
(\vec{r}_i-\vec{r}_j)^2
\right] \; .
\end{equation}
The one-body term of the Hamiltonian (\ref{hamomega}) can be re-written
as a sum of the center-of-mass term,
$H^\Omega_{\rm cm}=\frac{\vec{P}_{\rm cm}^2}{2Am}
+\frac{1}{2}Am\Omega^2 \vec{R}^2$, where
$\vec{P}_{\rm cm}=\sum_{i=1}^A \vec{p}_i$,
and a term depending only on the relative coordinates.
In the present application we use a basis, which explicitly separates
center-of-mass and relative-coordinate wave functions. Therefore,
the contribution of the center-of-mass term is trivial and will 
be omitted from now on. 

The shell-model calculations are performed in a finite model space. 
Therefore, the interaction term in Eq. (\ref{hamomega}) must be replaced
by an effective interaction. In general, for an $A$-nucleon system,
an $A$-body
effective interaction is needed. In practice, the effective
interaction is usually approximated by a two-body effective interaction.
In the present study we will also employ a three-body effective 
interaction. 
As approximations are involved in the effective interaction
treatment, large model spaces
are desirable. In that case, the calculation
should be less affected by any imprecision of the effective
interaction. 
The same is true for the evaluation of any observable characterized
by an operator. In the model space, renormalized effective operators 
are required. The larger the model space, the less renormalization
is needed.  
We may take advantage of the present approach
to perform shell-model calculations in significantly larger
model spaces than are possible in the conventional shell-model
approach.
At the same time we can investigate convergence properties
of effective interactions.

\subsection{Three-nucleon system}

In this subsection we repeat the steps discussed in Ref. \cite{NB98}
that are needed to solve the three-nucleon shell-model problem
in the Faddeev formalism.
For a three-nucleon system, i.e., $A=3$, 
the following transformation of the coordinates 
\begin{mathletters}\label{rtrans}
\begin{eqnarray}
\vec{r}&=&\sqrt{\textstyle{\frac{1}{2}}}(\vec{r}_1-\vec{r}_2) \; ,
\\
\vec{y}&=&\sqrt{\textstyle{\frac{2}{3}}}[\textstyle{\frac{1}{2}}(\vec{r}_1
+\vec{r}_2)-\vec{r}_3] \; ,
\end{eqnarray}
\end{mathletters}
and, similarly, of the momenta,
can be introduced that brings the relative-coordinate part of the
one-body HO Hamiltonian into the form
\begin{equation}\label{H0}
H_0 =  \frac{\vec{p}^2}{2m} + \frac{1}{2}m\Omega^2 \vec{r}^2
     + \frac{\vec{q}^2}{2m} + \frac{1}{2}m\Omega^2 \vec{y}^2 \; .
\end{equation}
Eigenstates of this Hamiltonian, 
\begin{equation}\label{hobas}
|n l s j t, {\cal N} {\cal L} {\cal J}, J T \rangle \; ,
\end{equation}
are then used as the basis for the three-nucleon calculation.
Here $n, l$ and ${\cal N}, {\cal L}$ are the HO quantum numbers
corresponding to the harmonic oscillators associated with the coordinates and 
momenta $\vec{r}, \vec{p}$ and $\vec{y}, \vec{q}$, respectively. 
The quantum numbers $s,t,j$ describe the spin, isospin and angular momentum
of the relative-coordinate partial channel of particles 1 and 2, while 
${\cal J}$ is the angular momentum of the third particle relative to the
center of mass of particles 1 and 2. The $J$ and $T$ are the total angular 
momentum and the total isospin, respectively.

The Faddeev equation for the bound system can be written in the form
\begin{equation}\label{Fadeq}
\tilde{H}|\phi\rangle = E|\phi\rangle \; ,
\end{equation}
with
\begin{equation}\label{Fadham}
\tilde{H}= H_0 + V(\vec{r}) {\cal T} \; .
\end{equation}
Here, 
$V(\vec{r})=V_{\rm N}(\sqrt{2}\vec{r})-\textstyle{\frac{1}{A}}m
\Omega^2 \vec{r}^2$
is the potential and ${\cal T}$ is given by
\begin{equation}\label{metric}
{\cal T}=1+{\cal T}^{(-)}+{\cal T}^{(+)} \; ,
\end{equation}
with ${\cal T}^{(+)}$ and ${\cal T}^{(-)}$ the cyclic and the anticyclic 
permutation 
operators, respectively. Previously \cite{NB98},
we derived a simple formula for the matrix elements of 
${\cal T}^{(-)}+{\cal T}^{(+)}$ in the basis (\ref{hobas}), namely
\begin{eqnarray}\label{t13t23}
&&\langle n_1 l_1 s_1 j_1 t_1, {\cal N}_1 {\cal L}_1  
{\cal J}_1, J T | {\cal T}^{(-)}+{\cal T}^{(+)} |  
n_2 l_2 s_2 j_2 t_2, {\cal N}_2 {\cal L}_2  
{\cal J}_2, J T\rangle = - \delta_{N_1,N_2} \nonumber \\
&& \times \sum_{LS} \hat{L}^2 \hat{S}^2
\hat{j}_1 \hat{j}_2 \hat{\cal J}_1 \hat{\cal J}_2 \hat{s}_1 \hat{s}_2
\hat{t}_1 \hat{t}_2 (-1)^L
          \left\{ \begin{array}{ccc} l_1   & s_1   & j_1   \\
          {\cal L}_1  & \textstyle{\frac{1}{2}}  & {\cal J}_1 \\
                                     L   & S  & J
\end{array}\right\}
          \left\{ \begin{array}{ccc} l_2   & s_2   & j_2   \\
       {\cal L}_2  & \textstyle{\frac{1}{2}}   & {\cal J}_2 \\
                                     L   & S  & J
\end{array}\right\}
  \left\{ \begin{array}{ccc} \textstyle{\frac{1}{2}} & \textstyle{\frac{1}{2}} 
               & s_1 \\
             \textstyle{\frac{1}{2}}  &  S  & s_2
\end{array}\right\}
  \left\{ \begin{array}{ccc} \textstyle{\frac{1}{2}} & \textstyle{\frac{1}{2}} 
               & t_1 \\
             \textstyle{\frac{1}{2}}  &  T  & t_2
\end{array}\right\}
\nonumber \\
&&\times \left[(-1)^{s_1+s_2+t_1+t_2-{\cal L}_1-l_1} 
\langle {\cal N}_1 {\cal L}_1 n_1 l_1 L 
| n_2 l_2 {\cal N}_2 {\cal L}_2 L \rangle_{\rm 3}
+\langle n_1 l_1 {\cal N}_1 {\cal L}_1 L 
| {\cal N}_2 {\cal L}_2 n_2 l_2 L \rangle_{\rm 3}
\right] \; ,
\end{eqnarray}
where $N_i=2n_i+l_i+2{\cal N}_i+{\cal L}_i, i\equiv 1,2$, 
$\hat{j}=\sqrt{2j+1}$ 
and $\langle {\cal N}_1 {\cal L}_1 n_1 l_1 L 
| n_2 l_2 {\cal N}_2 {\cal L}_2 L \rangle_{\rm 3}$
is the general HO bracket for two particles with mass 
ratio 3 as defined, e.g.,
in Ref. \cite{Tr72}.
The expression (\ref{t13t23}) can be derived by examining 
the action
of ${\cal T}^{(+)}$ and ${\cal T}^{(-)}$ on the basis states (\ref{hobas}).
A similar derivation for a different basis is described, e.g., 
in Refs. \cite{HKT72,Glo83}. Let us note that it follows 
from the antisymmetry of the two-nucleon states
and from the symmetry properties of the HO brackets
that the contributions of ${\cal T}^{(-)}$ and ${\cal T}^{(+)}$
in (\ref{t13t23}) are identical.

The eigensystem of the operator ${\cal T}$ (\ref{metric}) consists
of two subspaces. The first subspace has eigenstates with the eigenvalue
3, which form totally antisymmetric physical states. The second 
subspace has eigenstates with
the eigenvalue 0, which form a not completely antisymmetric, unphysical 
subspace of states. 
We found these properties of ${\cal T}$
by direct calculation using the relation (\ref{t13t23}). It is, however,
a general result. The same structure of eigenstates was also obtained
in Ref. \cite{RY95} using a different basis. The eigenvalue structure
follows from the fact that $\frac{1}{3}{\cal T}$ has the properies of
a projection operator.
It is possible to hermitize the 
Hamiltonian (\ref{Fadham}) on the physical subspace, where it is 
quasi-Hermitian.
The Hermitized Hamiltonian takes the form
\begin{equation}\label{Fadhamh}
\bar{H}= H_0 + \bar{{\cal T}}^{1/2}V(\vec{r})\bar{{\cal T}}^{1/2} \; ,
\end{equation}
where $\bar{{\cal T}}$ operates on the physical subspace only.

The operator ${\cal T}$ (\ref{metric}) is diagonal in 
$N=2n+l+2{\cal N}+{\cal L}$. Note that any basis truncation other 
than one of the type
$N\le N_{\rm max}$ violates, in general, the Pauli principle and mixes 
physical and unphysical states. 
Here, $N_{\rm max}$ characterizes the maximum of total allowed 
HO quanta in the model space and is an input parameter
of the calculation. 
The truncation into totally allowed
oscillator quanta $N\le N_{\rm max}$, however, 
preserves the equivalence of the
Hamiltonians (\ref{Fadham}) and (\ref{Fadhamh}) on the physical subspace.

\subsection{Four-nucleon system}

By relying on the results obtained for the three-nucleon system,
as described in the previous subsection, we can extend
the formalism to the four-nucleon system. 
We use the Hamiltonian (\ref{hamomega}) with $A=4$. By introducing 
the coordinate (and momentum) transformations,
\begin{mathletters}\label{fourtran}
\begin{eqnarray}
\vec{r}&=&\sqrt{\textstyle{\frac{1}{2}}}(\vec{r}_1-\vec{r}_2) \;, 
\\
\vec{y}&=&\sqrt{\textstyle{\frac{2}{3}}}[\textstyle{\frac{1}{2}}(\vec{r}_1
+\vec{r}_2)-\vec{r}_3] \;, 
\\
\vec{z}&=&\textstyle{\frac{\sqrt{3}}{2}}[\textstyle{\frac{1}{3}}(\vec{r}_1
+\vec{r}_2+\vec{r}_3)-\vec{r}_4] \;, 
\end{eqnarray}
\end{mathletters}
we obtain the one-body part 
of the Hamiltonian (\ref{hamomega}) in the form
\begin{equation}
H_0 =  \frac{\vec{p}^2}{2m} + \frac{1}{2}m\Omega^2 \vec{r}^2
     + \frac{\vec{q}^2}{2m} + \frac{1}{2}m\Omega^2 \vec{y}^2 
     + \frac{\vec{o}^2}{2m} + \frac{1}{2}m\Omega^2 \vec{z}^2 \; ,
\end{equation}
with the center-of-mass term omitted.

A possible generalization of the Faddeev equation (\ref{Fadeq})
for four identical particles can be written in the form  
\begin{equation}\label{fadfour}
\tilde{H}|\psi_{(123)4}\rangle = E|\psi_{(123)4}\rangle \; ,
\end{equation}
with
\begin{equation}
\tilde{H}|\psi_{(123)4}\rangle \equiv  H_0 |\psi_{(123)4}\rangle 
+ \textstyle{\frac{1}{2}}(V_{12}+V_{13}+V_{23})
(|\psi_{(123)4}\rangle 
+|\psi_{(432)1}\rangle +|\psi_{(134)2}\rangle 
+|\psi_{(142)3}\rangle )\; ,
\end{equation}
and
\begin{equation}\label{metric4}
(|\psi_{(123)4}\rangle 
+|\psi_{(432)1}\rangle +|\psi_{(134)2}\rangle 
+|\psi_{(142)3}\rangle ) =
(1-{\cal T}_{14}-{\cal T}_{24}-{\cal T}_{34})|\psi_{(123)4}\rangle 
\equiv {\cal T}_4 |\psi_{(123)4}\rangle \; .
\end{equation}
Here, $|\psi_{(123)4}\rangle $ is a four-fermion Faddeev amplitude
completely antisymmetrized for particles 1,2, and 3. There are 
three other equations that can be obtained from Eq. (\ref{fadfour}) by
permuting particle 4 with particles 1, 2, and 3. Their sum then leads
to the Schr\"odinger equation. We note that the present equations
are different from the traditional Faddeev-Yakubovsky equations
\cite{Ya67}, which combine Faddeev amplitudes depending on two
sets of relative coordinates. We are working with a complete orthonormal
basis. It is, therefore, sufficient and convenient to use a single set
of coordinates defined by the relations (\ref{fourtran}).
Unlike the Faddeev amplitudes used typically in the Faddeev-Yakubovsky
equations, the amplitudes appearing in Eq. (\ref{fadfour}) are 
antisymmetrized with respect to the first three particles.
Those amplitudes are obtained, as described below, in a straightforward
manner with the help of our three-nucleon HO formalism
introduced earlier.
The present equations allow us to employ easily real three-body interactions
or three-body effective interactions. The latter property makes 
them particularly useful for the present extension of shell-model 
calculations for four nucleons. At the same time, the use of Faddeev 
amplitudes antisymmetrized for particles 1, 2 and 3 allows us
to reduce the dimmension of the basis significantly.

We start the four-nucleon calculation
using the basis
\begin{equation}\label{fourbas}
|N_1 i J_1 T_1, n_z l_z {\cal J}_4, J T \rangle \; .
\end{equation}
with the three-fermion part given by
the antisymmetrized eigenstates of ${\cal T}$ (\ref{metric}) 
corresponding to eigenvalue 3, e.g., 
\begin{equation}\label{threebant}
|N_1 i J_1 T_1\rangle =
\sum c_{n l s j t {\cal N} {\cal L}
{\cal J}_{3}}^{N_{1 } i J_{ 1} T_{ 1}}
|n l s j t, {\cal N} {\cal L} {\cal J}_3, J_1 T_1\rangle
\; ,
\end{equation}
where $N_1=2n+l+2{\cal N}+{\cal L}$ and $i$ counts the eigenstates
of ${\cal T}$ with the eigenvalue 3 for given $N_1$ and $J_1,T_1$.
Further,  $n_z, l_z$ are the HO quantum numbers
corresponding to the harmonic oscillator associated with 
the coordinate $\vec{z}$ and the momentum $\vec{o}$ and 
${\cal J}_4$ is the angular momentum of the fourth particle 
relative to the center of mass of particles 1, 2 and 3.

As in the case of the three-particle transposition operators
(\ref{t13t23}),
a compact formula can be derived for the matrix elements
of the four-particle transposition operators in the basis
(\ref{fourbas}), e.g.,
\begin{eqnarray}\label{t4}
&&\langle N_{\rm 1L} i_{\rm L} J_{\rm 1L} T_{\rm 1L}, n_{z \rm L}
l_{z \rm L} {\cal J}_{\rm 4L}, JT | {\cal T}_{14}
+{\cal T}_{24}+{\cal T}_{34}
|N_{\rm 1R} i_{\rm R} J_{\rm 1R} T_{\rm 1R}, n_{z \rm R}
l_{z \rm R} {\cal J}_{\rm 4R}, JT \rangle \nonumber \\
&&=\delta_{N_{\rm L},N_{\rm R}} \sum
c_{n_{\rm L} l_{\rm L} s_{\rm L} j_{\rm L} t_{\rm L} 
{\cal N}_{\rm L} {\cal L}_{\rm L}
{\cal J}_{3\rm L}}^{N_{1 \rm L} i_{\rm L} J_{\rm 1L} T_{\rm 1L}}
c_{n_{\rm R} l_{\rm R} s_{\rm R} j_{\rm R} t_{\rm R} 
{\cal N}_{\rm R} {\cal L}_{\rm R}
{\cal J}_{3\rm R}}^{N_{1 \rm R} i_{\rm R} J_{\rm 1R} T_{\rm 1R}}
\hat{L}_{\rm 1L}^2 \hat{L}_{\rm 1R}^2 \hat{S}_{\rm 1L}^2
\hat{S}_{\rm 1R}^2 \hat{L}_2^2 \hat{S}_2^2
\nonumber \\
&&\times \hat{j}_{\rm L} \hat{j}_{\rm R} \hat{\cal J}_{\rm 3L} 
\hat{\cal J}_{\rm 3R} \hat{\cal J}_{\rm 4L} 
\hat{\cal J}_{\rm 4R}\hat{J}_{\rm 1L}\hat{J}_{\rm 1R}
\hat{T}_{\rm 1L}\hat{T}_{\rm 1R}
(-1)^{T_{\rm 1L}-T_{\rm 1R}+S_{\rm 1L}+S_{\rm 1R}}
\left\{ \begin{array}{ccc} \textstyle{\frac{1}{2}} & s_{\rm R} 
               & S_{\rm 1R} \\
             \textstyle{\frac{1}{2}}  &  S_2  & S_{\rm 1L}
\end{array}\right\}
\left\{ \begin{array}{ccc} \textstyle{\frac{1}{2}} & t_{\rm R} 
               & T_{\rm 1R} \\
             \textstyle{\frac{1}{2}}  &  T  & T_{\rm 1L}
\end{array}\right\}
\nonumber \\
&&\times 
\left\{ \begin{array}{ccc} l_{\rm L}   & s_{\rm L}   & j_{\rm L}   \\
  {\cal L}_{\rm L}  & \textstyle{\frac{1}{2}}  & {\cal J}_{\rm 3L} \\
                           L_{\rm 1L}   & S_{\rm 1L}  & J_{\rm 1L}
\end{array}\right\}
\left\{ \begin{array}{ccc} l_{\rm R}   & s_{\rm R}   & j_{\rm R}   \\
  {\cal L}_{\rm R}  & \textstyle{\frac{1}{2}}  & {\cal J}_{\rm 3R} \\
                           L_{\rm 1R}   & S_{\rm 1R}  & J_{\rm 1R}
\end{array}\right\}
\left\{ \begin{array}{ccc} L_{\rm 1L}   & S_{\rm 1L}   & J_{\rm 1L}   \\
  l_{z\rm L}  & \textstyle{\frac{1}{2}}  & {\cal J}_{\rm 4L} \\
                           L_2   & S_2  & J
\end{array}\right\}
\left\{ \begin{array}{ccc} L_{\rm 1R}   & S_{\rm 1R}   & J_{\rm 1R}   \\
  l_{z\rm R}  & \textstyle{\frac{1}{2}}  & {\cal J}_{\rm 4R} \\
                           L_2   & S_2  & J
\end{array}\right\}
\nonumber \\
&&\times \hat{L}'^2 (-1)^{L'}
\left\{ \begin{array}{ccc} l_{\rm R} & L_2 & L' \\
             l_{z\rm R}  &  {\cal L}_{\rm R}  & L_{\rm 1R}
\end{array}\right\}
\left\{ \begin{array}{ccc} l_{\rm R} & L_2 & L' \\
             l_{z\rm L}  &  l'  & L_{\rm 1L}
\end{array}\right\}
\left[
\hat{s}_{\rm L} \hat{s}_{\rm R}\hat{t}_{\rm L} \hat{t}_{\rm R}
  \left\{ \begin{array}{ccc} \textstyle{\frac{1}{2}} & \textstyle{\frac{1}{2}} 
               & s_{\rm R} \\
             \textstyle{\frac{1}{2}}  &  S_{\rm 1L}  & s_{\rm L}
\end{array}\right\}
  \left\{ \begin{array}{ccc} \textstyle{\frac{1}{2}} & \textstyle{\frac{1}{2}} 
               & t_{\rm R} \\
             \textstyle{\frac{1}{2}}  &  T_{\rm 1L}  & t_{\rm L}
\end{array}\right\}
\right.
\nonumber \\
&&\times (-1)^{l_{z\rm R}+L_{\rm 1L}}
\left( (-1)^{l_{z\rm L}}
\langle n' l' n_{z\rm L} l_{z\rm L} L' 
| n_{z\rm R} l_{z\rm R} {\cal N}_{\rm R} {\cal L}_{\rm R} L' 
\rangle_{\rm 8} 
\langle n_{\rm L} l_{\rm L} {\cal N}_{\rm L} {\cal L}_{\rm L} L_{\rm 1L} 
| n' l'  n_{\rm R} l_{\rm R} L_{\rm 1L} 
\rangle_{\rm 3} \right.
\nonumber \\
&&\left. 
+(-1)^{t_{\rm R}-t_{\rm L}+s_{\rm R}-s_{\rm L}
+{\cal L}_{\rm R}
-l_{\rm L}-{\cal L}_{\rm L}}
\langle n_{z\rm L} l_{z\rm L} n' l' L' 
| {\cal N}_{\rm R} {\cal L}_{\rm R} n_{z\rm R} l_{z\rm R}  L' 
\rangle_{\rm 8} 
\langle {\cal N}_{\rm L} {\cal L}_{\rm L} n_{\rm L} l_{\rm L} L_{\rm 1L} 
| n_{\rm R} l_{\rm R} n' l' L_{\rm 1L} 
\rangle_{\rm 3}\right)
\nonumber \\
&&\left. - \delta_{l_{\rm L},l_{\rm R}}\delta_{s_{\rm L},s_{\rm R}}
\delta_{t_{\rm L},t_{\rm R}}\delta_{{\cal N}_{\rm L},n'}
\delta_{{\cal L}_{\rm L},l'} (-1)^{{\cal L}_{\rm R}+l_{z\rm R}}
\langle n_{z\rm L} l_{z\rm L}  {\cal N}_{\rm L} {\cal L}_{\rm L} L' 
| {\cal N}_{\rm R} {\cal L}_{\rm R} n_{z\rm R} l_{z\rm R}  L' 
\rangle_{\rm 8} \right] \; ,
\end{eqnarray}
where 
$N_{\rm X}=2n_{\rm X}+l_{\rm X}+2{\cal N}_{\rm X}+{\cal L}_{\rm X}
+2n_{z\rm X}+l_{z\rm X}, {\rm X}\equiv {\rm L \; or \; R}$,
and, e.g., the expression
$\langle n_{z\rm L} l_{z\rm L}  {\cal N}_{\rm L} {\cal L}_{\rm L} L' 
| {\cal N}_{\rm R} {\cal L}_{\rm R} n_{z\rm R} l_{z\rm R}  L' 
\rangle_{\rm 8}$ denotes a general HO bracket 
for two particles with mass ratio 8, as defined
in Ref. \cite{Tr72}. Similarly, as in Eq. (\ref{t13t23}) the brackets
for two particles with mass ratio 3 also appear 
in the relation (\ref{t4}). In the derivation of the expression
(\ref{t4}) we relied on the antisymmetry of the basis states with respect
to particles 1,2 and 3. The calculation was facilitated by application
of the operators $-{\cal T}_{13}$ and $-{\cal T}_{23}$. The relation
(\ref{t4}) appears to be non-symmetric. However, its numerical evaluation
leads to a symmetric matrix. It may also appear that the angular momentum
sums in (\ref{t4}) can be summed up. In fact, it is possible to
simplify the expression by introducing a $15j$-coefficient of the
fifth kind as defined, e.g., in Ref. \cite{YLV60}, but, as such
coefficients are seldomly used, we prefer to keep the summations
in the explicit form. On the other hand, a significant simplification
of the expression (\ref{t4}) can be obtained, when the symmetry relations
of different terms are exploited. First, it follows from the properties
of the HO brackets and from the antisymmetry of the two-nucleon
states that the contributions of ${\cal T}_{14}$ and ${\cal T}_{24}$
are identical. Second, using the fact that the states (\ref{fourbas})
are antisymmetrized for the particles 1, 2 and 3 it follows that
all three permutation operators appearing in (\ref{t4}) give identical
contributions to the expression (\ref{t4}). 
The computation of ${\cal T}_{34}$ is the simplest. In that case
a partial summation of the angular momentum coefficients can be performed,
yielding a compact expression
\begin{eqnarray}\label{t34}
&&\langle N_{\rm 1L} i_{\rm L} J_{\rm 1L} T_{\rm 1L}, n_{z \rm L}
l_{z \rm L} {\cal J}_{\rm 4L}, JT | {\cal T}_{34}
|N_{\rm 1R} i_{\rm R} J_{\rm 1R} T_{\rm 1R}, n_{z \rm R}
l_{z \rm R} {\cal J}_{\rm 4R}, JT \rangle \nonumber \\
&&=\delta_{N_{\rm L},N_{\rm R}} \sum
c_{n l s j t 
{\cal N}_{\rm L} {\cal L}_{\rm L}
{\cal J}_{3\rm L}}^{N_{1 \rm L} i_{\rm L} J_{\rm 1L} T_{\rm 1L}}
c_{n l s j t 
{\cal N}_{\rm R} {\cal L}_{\rm R}
{\cal J}_{3\rm R}}^{N_{1 \rm R} i_{\rm R} J_{\rm 1R} T_{\rm 1R}}
\nonumber \\
&&\times \hat{\cal J}_{\rm 3L} 
\hat{\cal J}_{\rm 3R} \hat{\cal J}_{\rm 4L} 
\hat{\cal J}_{\rm 4R}\hat{J}_{\rm 1L}\hat{J}_{\rm 1R}
\hat{T}_{\rm 1L}\hat{T}_{\rm 1R}
(-1)^{T_{\rm 1L}+T_{\rm 1R}+{\cal J}_{\rm 3L}+{\cal J}_{\rm 3R}}
\left\{ \begin{array}{ccc} \textstyle{\frac{1}{2}} & t 
               & T_{\rm 1R} \\
             \textstyle{\frac{1}{2}}  &  T  & T_{\rm 1L}
\end{array}\right\}
\nonumber \\
&&\times 
\hat{K}^2
\left\{ \begin{array}{ccc} j   & {\cal J}_{\rm 3L}    & J_{\rm 1L}   \\
  {\cal J}_{\rm 3R}  & K  & {\cal J}_{\rm 4L} \\
                           J_{\rm 1R}   & {\cal J}_{\rm 4R}  & J
\end{array}\right\}
\left\{ \begin{array}{ccc} {\cal L}_{\rm L}   & l_{\rm zR}   & K   \\
  {\cal J}_{\rm 4R}  & {\cal J}_{\rm 3L}& \textstyle{\frac{1}{2}}   
\end{array}\right\}
\left\{ \begin{array}{ccc} {\cal L}_{\rm R}   & l_{\rm zL}   & K   \\
  {\cal J}_{\rm 4L}  & {\cal J}_{\rm 3R}& \textstyle{\frac{1}{2}}   
\end{array}\right\}
\left\{ \begin{array}{ccc} l_{\rm zL}   & {\cal L}_{\rm R}   &  K   \\
                           l_{\rm zR}   & {\cal L}_{\rm L}   &  L'
\end{array}\right\}
\nonumber \\
&&\times 
\hat{L}'^2 (-1)^{{\cal L}_{\rm R}+l_{z\rm L}+L'}
\langle n_{z\rm L} l_{z\rm L}  {\cal N}_{\rm L} {\cal L}_{\rm L} L' 
| {\cal N}_{\rm R} {\cal L}_{\rm R} n_{z\rm R} l_{z\rm R}  L' 
\rangle_{\rm 8}  \; .
\end{eqnarray}
Thus, by multiplying the expression (\ref{t34}) by three we obtain
the same matrix element as from (\ref{t4}). We note that a generalization
of the evaluation of the permutation operator matrix element (\ref{t34})
to a more
complex system, than the presently studied $A=4$, is straightforward.
Its simplicity suggests that the present formalism can be extended
to systems with $A>4$.

Similarly, as for the operator ${\cal T}$ (\ref{metric}),
eigenstates of the operator ${\cal T}_4$ defined 
by the relation (\ref{metric4}) can be subdivided into two subspaces.
A physical subspace is spanned by totally antisymmetric states, 
in this case corresponding to the eigenvalue 4, 
and a spurious subspace is spanned by eigenvectors
corresponding to the eigenvalue 0.
It is possible to symmetrize the Hamiltonian $\tilde{H}$ appearing in 
Eq. (\ref{fadfour}) on the
physical subspace. The symmetrized Hamiltonian then takes the form  
\begin{equation}\label{foureq}
\bar{H}= H_0 + \bar{{\cal T}_4}^{1/2}\textstyle{\frac{1}{2}}
(V_{12}+V_{13}+V_{23})\bar{{\cal T}_4}^{1/2} \; ,
\end{equation}
where $\bar{\cal T}_4$ operates only on the physical subspace.
In our calculations, described later, we diagonalize 
the symmetrized Hamiltonian (\ref{foureq}) in the physical basis 
formed by the eigenstates of $\bar{\cal T}_4$. 

The operator ${\cal T}_4$ (\ref{metric4}) is diagonal in 
$N=2n+l+2{\cal N}+{\cal L}+2n_z+l_z$. 
A basis truncation defined by a restriction on the totally 
allowed oscillator quanta $N\le N_{\rm max}$ preserves 
the equivalence of the Hamiltonians (\ref{Fadham}) and 
(\ref{Fadhamh}) on the physical subspace.

\subsection{Effective interactions}

From solving two-nucleon systems 
in a HO well, interacting by soft-core potentials,
one learns that excitations up to about $300\hbar\Omega$ ($N_{\rm max}=300$)
are required to get almost exact solutions. We anticipate,
therefore, that at least the same number of excitations should
be allowed to solve the many-nucleon system.
The Faddeev formulation
has the obvious advantage compared with the traditional shell-model
approach that the center-of-mass coordinate
is explicitly removed. Even then, it is not
feasible to solve the eigenvalue problem either for (\ref{Fadhamh})
or for (\ref{foureq}) in such a large space.
On the other hand, shell-model calculations are always performed by
employing effective interactions tailored to a specific
model space. In practice, these effective interactions can
never be calculated exactly, because, in general, for  
an A-body effective interaction is required for an A-nucleon system. 
We may, however, take advantage of the present approach
to perform shell-model calculations in significantly larger
model spaces than are possible in conventional shell-model
approach.
At the same time we can investigate convergence properties
of effective interactions. If convergence is achieved,
we should obtain the exact solution, since by construction
the effective interactions that we employ satisfy 
the condition $V_{\rm eff}\rightarrow V$ for 
$N_{\rm max}\rightarrow\infty$.

Usually, the effective interaction is approximated by a 
two-body effective interaction determined from a two-nucleon
system.    
In the present calculations we replace matrix elements of the potential
$V(\vec{r})$ by matrix elements of an effective two-body
interaction, derived in a straightforward manner for each
relative-coordinate partial channel. 
The relevant two-nucleon Hamiltonian is then
\begin{equation}\label{hamomega2}
H_2\equiv H_{02}+V=
\frac{\vec{p}^2}{2m}
+\frac{1}{2}m\Omega^2 \vec{r}^2
+ V_{\rm N}(\sqrt{2}\vec{r})-\frac{m\Omega^2}{A}\vec{r}^2 \; ,
\end{equation}
which can be solved as a differential equation or, alternatively,
can be diagonalized in a sufficiently large harmonic oscillator
basis. For a four-nucleon system we set $A=4$ in Eq. (\ref{hamomega2}),
which implies that we are dealing with a bound-state problem.

To construct the two-body effective interaction we employ
the Lee-Suzuki \cite{LS80} similarity transformation
method, which gives the effective interaction in the form
$P V_{\rm eff}P = P V P + PV Q\omega P$,
with $\omega$ the transformation operator satisfying $\omega=Q \omega P$,
and $P$ and $Q=1-P$, the projectors on the model and the complementary 
spaces, respectively. 
Our calculations start with exact solutions of the Hamiltonian
(\ref{hamomega2}) and, consequently, we construct
the operator $\omega$ and, then, the effective interaction directly
from these solutions. Let us denote the relative-coordinate two-nucleon 
HO states, which form the model space, 
as $|\alpha_P\rangle$,
and those which belong to the Q-space, as $|\alpha_Q\rangle$.
Then the Q-space components of the eigenvector $|k\rangle$ of
the Hamiltonian (\ref{hamomega2}) can be expressed as a combination
of the P-space components with the help of the operator $\omega$
\begin{equation}\label{eigomega}  
\langle\alpha_Q|k\rangle=\sum_{\alpha_P}
\langle\alpha_Q|\omega|\alpha_P\rangle \langle\alpha_P|k\rangle \; .
\end{equation}
If the dimension of the model space is $d_P$, we may choose a set
${\cal K}$ of $d_P$ eigenevectors, 
for which the relation (\ref{eigomega}) 
will be satisfied. Under the condition that the $d_P\times d_P$ 
matrix $\langle\alpha_P|k\rangle$ for $|k\rangle\in{\cal K}$
is invertible, the operator $\omega$ can be determined from 
(\ref{eigomega}).  In the present application we select the lowest states
obtained in each channel. Their number is given by the number of basis
states satisfying $2n+l\leq N_{\rm max}$. 
Once the operator $\omega$ is determined, the effective hamiltonian
can be constructed as follows 
\begin{equation}\label{effomega}
\langle \gamma_P|H_{2\rm eff}|\alpha_P\rangle =\sum_{k\in{\cal K}}
\left[
\langle\gamma_P|k\rangle E_k\langle k|\alpha_P\rangle
+\sum_{\alpha_Q}\langle\gamma_P|k\rangle E_k\langle k|\alpha_Q\rangle
\langle\alpha_Q |\omega|\alpha_P\rangle\right] \; .
\end{equation}
It should be noted that 
$P|k\rangle=\sum_{\alpha_P}|\alpha_P\rangle\langle\alpha_P|k\rangle$
for $|k\rangle\in{\cal K}$ is a right eigenvector of (\ref{effomega})
with the eigenvalue $E_k$.

This Hamiltonian, when diagonalized in a model-space basis, reproduces
exactly the set ${\cal K}$ of $d_P$ eigenvalues $E_k$. Note that
the effective Hamiltonian is, in general, quasi-Hermitian. 
It can be hermitized by a similarity transformation 
determined from the metric operator $P(1+\omega^\dagger\omega)P$. 
The Hermitian Hamiltonian is then given by \cite{S82SO83}
\begin{equation}\label{hermeffomega}
\bar{H}_{\rm 2eff}
=\left[P(1+\omega^\dagger\omega)P\right]^{1/2}
H_{\rm 2eff}\left[P(1+\omega^\dagger\omega)
P\right]^{-1/2} \; .
\end{equation}

Finally, the two-body effective interaction used 
in the present calculations
is determined from the two-nucleon effective Hamiltonian 
(\ref{hermeffomega}) as $V_{\rm 2eff}=\bar{H}_{\rm 2eff}-H_{02}$.
We note that the interaction $V_{12}+V_{13}+V_{23}$ in Eq. (\ref{foureq})
is then replaced by ${\cal T}^{1/2}V_{\rm 2eff}{\cal T}^{1/2}$,
which is evaluated in a straightforward way in the basis (\ref{fourbas}).

As pointed out before, the structure of the Hamiltonian (\ref{foureq})
allows us to employ easily three-body effective interactions in addition to
the above discussed two-body effective interactions. We can replace
$V_{12}+V_{13}+V_{23}$ in Eq. (\ref{foureq}) by $V_{\rm 3eff}$ that
can be derived from the three-nucleon solutions in a similar manner
as the two-body effective interaction is derived from the two-nucleon 
solutions. To find $V_{\rm 3eff}$ we solve the three-nucleon system
described by the Hamiltonian (\ref{Fadhamh}) with 
$V(\vec{r})=V_{\rm N}(\sqrt{2}\vec{r})-\textstyle{\frac{1}{A}}m
\Omega^2 \vec{r}^2$. As $A=4$ we are dealing with a bound three-nucleon
problem. It can be solved in a three-nucleon model space characterized
by $N_{\rm 3max}\approx 30$ \cite{NB98}. 
First, we compute the two-body effective 
interaction appropriate for the model space defined by $N_{\rm 3max}$,
as discussed earlier in this subsection.
Then the three-nucleon system is solved in the same space. Afterwards
we construct the three-body effective interaction for a model space
defined by $N_{\rm max}<N_{\rm 3max}$. In the present paper we use model
spaces up to $N_{\rm max}=14$. The effective interaction is constructed
exactly, as described above, using Eqs. 
(\ref{eigomega},\ref{effomega},\ref{hermeffomega}) with $H_{\rm 2eff}$
replaced by $H_{\rm 3eff}$. 
The energies $E_k$ and the states $|k\rangle$ correspond
to the three-nucleon system eigenstates, however, and the states
$|\alpha_P\rangle$ and $|\alpha_Q\rangle$ are three-nucleon basis
states (\ref{threebant}) with the model-space condition 
$N_1\equiv 2n+l+2{\cal N}+{\cal L}\leq N_{\rm max}$ 
and the $Q$-space condition
$N_{\rm max}<N_1\leq N_{\rm 3max}$. The three-body effective interaction
is computed for different three-nucleon channels characterized by
$J_1, T_1$ and parity and is obtained from the hermitized effective
Hamiltonian as $V_{\rm 3eff}=\bar{H}_{\rm 3eff}-H_{0}$, where
$H_0$ is given by Eq. (\ref{H0}). The interaction $V_{\rm 3eff}$ then
replaces $V_{12}+V_{13}+V_{23}$ in Eq. (\ref{foureq}).
We note that by construction in the limit 
$N_{\rm max}\rightarrow N_{\rm 3max}$ the three-body effective interaction
approaches the two-body effective interaction 
$V_{\rm 3eff}\rightarrow {\cal T}^{1/2}V_{\rm 2eff}{\cal T}^{1/2}$ 
and with $N_{\rm max}\rightarrow \infty$ the effective interaction
approaches the bare interaction $V_{\rm 2eff}\rightarrow V$.

\section{Application to $^4$He}
\label{sec3}

In the present paper we use the Reid93 NN
potential \cite{SKTS} and the Argonne V8' NN
potential, introduced in Ref. \cite{GFMC}.
We work in the isospin formalism; the charge invariant potential 
$V_{\rm N} =\frac{1}{3} V_{pp}+\frac{1}{3} V_{nn}+\frac{1}{3} V_{np}$
is used for each $T=1$ wave in the calculations with the Reid93 potential.
The Coulomb potential is added to $V_{pp}$ in this case. On the other
hand, the calculations with the Argonne V8' potential, which is
isopin invariant, do not include the Coulomb potential.

Our calculation progresses in several steps.
The model space is characterized by the condition $N\le N_{\rm max}$,
$N=2n+l+2{\cal N}+{\cal L}+2n_z+l_z$. First, the three-nucleon 
antisymmetrized basis is constructed by diagonalizing ${\cal T}$
(\ref{metric}) in the basis (\ref{hobas}) for all 
$N_1\equiv 2n+l+2{\cal N}+{\cal L}\le N_{\rm max}$ and all $J_1, T_1$.
Then the four-nucleon antisymmetrized basis is calculated
by diagonalizing ${\cal T}_4$ (\ref{metric4}) in the basis (\ref{fourbas})
for $N=N_1+2n_z+l_z\le N_{\rm max}$ with $N$ even for positive
parity states and $N$ odd for negative parity states. We present results
for $J=0$ and $T=0$ only, but for both parities. We note that the
four-nucleon basis computation is independent of $\Omega$ and is done
only once.  The next step is the effective interaction calculation. 
The two-body effective interaction is derived from the Eqs. 
(\ref{eigomega})-(\ref{hermeffomega}). The condition for 
the relative-coordinate two-body
effective-interaction model space is then $2n+l\le N_{\rm max}$.
When solving the two-nucleon relative-coordinate Hamiltonian 
(\ref{hamomega2}) in the full space, we truncate the HO 
basis by keeping the states typically up to $n = 152$. 
The two-body effective interaction is constructed for all 
partial-wave channels up to $j=6$. The resulting effective interaction
is finally used as input for the four-nucleon calculation,
where the Hamiltonian (\ref{foureq}) is diagonalized. 
Instead of a two-body effective interaction, we may use a three-body
effective interaction, as discussed in the previous section.
The three-body effective is computed only for the most important 
three-nucleon channels $J_1 T_1$. In particular, we evaluated
the three-body effective interaction for $J_1=1/2, 3/2$,
$T_1=1/2$ and for both positive and negative parity. For the channels with
higher $J_1$ the two-body effective interaction corresponding
to $N_{\rm max}$ is used instead. For the parameter $N_{\rm 3max}$
characterizing the three-nucleon full space, we used $N_{\rm 3max}=28$
for $J_1=1/2$ and $N_{\rm 3max}=24$ for $J_1=3/2$.
We also performed calculations with the inclusion of the 
three-body effective interaction for $J_1=5/2$ and found it to have 
little effect.

Let us remark that the present method for solving the four-nucleon
shell-model problem is fully equivalent to the standard shell-model
approach. In particular, it is straightforward to transform the 
relative-coordinate two-body effective interaction used in the present 
calculations to the two-particle basis used for the shell-model
input by the standard transformation \cite{HG83}. We used
the transformed interactions for the model spaces up to an 
8$\hbar\Omega$ space to test our results.
The shell-model diagonalization was then performed by employing the
Many-Fermion-Dynamics Shell-Model Code \cite{VZ94},
which can be utilized for calculations with model spaces comprising
up to 9 major HO shells, i.e., $N_{\rm max}=8$
for $^4$He.
We obtained the same results from both the Faddev-like calculation and
the standard shell-model calculation. The Faddeev-like calculation has,
obviously, much smaller dimension and can be extended to larger model 
spaces. We also note that we applied the discussed formalism  
to four-electron system in a related study recently \cite{NBG98}.
Our results compared well with those obtained by the Stochastic 
Variational Method \cite{VOS}.

\subsection{Energies and point-nucleon rms radii}

Our results for the ground-state and excited-state energies 
and point-nucleon rms radii
are presented in Figs. \ref{figr93e}-\ref{figv8p23}, where
the dependencies on the model-space size and the HO
energy are shown. A summary of the largest model-space ($N_{\rm max}=14$
for the positive-parity states and $N_{\rm max}=13$
for the negative-parity states)
results is given in Table \ref{tab1}. 
Let us mention an unusual feature
of the present calculations, namely, the convergence from below
for the ground-state energy. It is caused by the asymmetric treatment 
of the HO terms that are added and subtracted
to the Hamiltonian in the process of evaluating the effective
interaction. Our effective interaction
is computed for a two- or three-nucleon system bound in an 
HO potential. Therefore, artificial binding 
from this potential is included
in the effective interaction and the four-body effects coming from 
the entire four-nucleon calculation may not completely compensate 
for this spurious binding in a particular model space.
We note that this type of over-binding in the no-core shell-model 
calculations was noticed in previous studies \cite{ZBVC,ZBVM,NB96}. 
This effect decreases as the model-space size increases, 
as is demonstrated in our earlier three-nucleon shell-model
calculations \cite{NB98}.

In Fig. \ref{figr93e} we present the calculated dependence of the
ground-state energy and the first-excited $0^+ 0$ state energy
on the model-space size, characterized by $N_{\rm max}$. 
The two-body effective interaction employed was derived
from the Reid93 NN potential. 
Results for $\hbar\Omega=14, 17, 19, 22$ MeV are shown.
The corresponding dependence of the point-nucleon rms radius
is presented in Fig. \ref{figr93r}. A slow convergence with the
increasing model-space size can be observed for the energies
with a significantly faster rate for the ground state compared
to the first excited $0^+ 0$ state. Also, much stronger dependence
of the excited state on the HO
energy $\hbar\Omega$ is apparent. The results of the point-nucleon
rms-radius calculation demonstrate even more the differences
between the ground state and the first excited $0^+ 0$ state. 
While the ground-state
radius has almost converged and shows little $\hbar\Omega$ dependence,
the first excited $0^+ 0$ state displays a strong 
dependence of its energy on $\hbar\Omega$ and a steady 
increase of its radius with increasing model-space size.

Let us remark that in our approach we obtain the ground state
as well as the excited states by diagonalizing the Hamiltonian.
This implies that the excited states are expanded
in the same harmonic-oscillator basis used for the ground state.
While such an approach has technical advantages, it might not be 
physically sound. Cautious interpretation of the excited-state
results is, therefore, necessary.  
The significantly different convergence rate of the ground state
and of the first excited $0^+ 0$ state manifests the different 
nature of the two states.
Let us note that if the model-space size was increased up to the point
at which total convergence of the excited state was achieved, 
our procedure would
yield isolated three- and one-body clusters with an infinite rms radius
and a total energy of the three-nucleon system. It is possible, though,
that we could observe a meta-stability prior to the onset of the cluster
separation, as the resonance is sharp and low-lying. The present model-space
sizes, however, are not yet sufficient to arrive at that point.
That we have not reached this point
can be seen from the lack of convergence and, in particular,
from the rather small rms radius, which shows a significant increase
with $N_{\rm max}$ and a strong dependence on $\Omega$.    
 
The importance of the three-body effective interaction can be judged
from the results shown in Fig. \ref{figr93et}. The ground-state
and excited $0^+ 0$ state energies obtained in a calculation
that employs the three-body effective interaction is compared
to a calculation performed by using only the two-body effective 
interaction. Results for two different values of the HO energy,
$\hbar\Omega=17$ and 19 MeV, are presented. The dashed lines
connect the two-body effective interaction calculation results
that are identical to those in Fig. \ref{figr93e}
that correspond to HO energies of
$\hbar\Omega=17$ and 19 MeV. The full lines connect the 
results obtained in calculations with the three-body
effective interaction. It is apparent that the three-body effective
interaction improves the converegence considerably. It is especially
true for the ground state. The difference between the $N_{\rm max}=6$
and $N_{\rm max}=14$ energies is significantly smaller in the
calculation that employs the three-body effective interaction.
It can also be seen that the two-body effective interaction results
approach the three-body effective interaction results in the largest
spaces used in our calculations. In addition, the dependence on the HO 
energy decreases in the three-body effective interaction calculation
compared to the two-body effective interaction calculation. This
holds for both the ground state and the first excited $0^+ 0$ state. 
However, the inclusion of the three-body effective interaction clearly
has a larger overall impact on the ground-state results.
  
The influence of the three-body effective interaction on 
the point-nucleon rms radius is depicted in Fig. \ref{figr93rt}.
Again we observe a better stability of the radii computed
using the three-body effective interaction. In particular,
the ground-state point-nucleon rms radius shows convergence
in both the model-space-size dependence and the 
HO-frequency dependence. On the other hand, the three-body effective 
interaction does not improve the convergence of the excited state
in any significant way in the model spaces that we employed.

In Fig. \ref{figr93etpm} we show the calculated energies of the
first $0^- 0$ state obtained using two-body effective
interactions in model spaces up to $N_{\rm max}=13$. For a comparison,
the results for the ground state and the first excited $0^+ 0$ state
from Fig. \ref{figr93et} are also presented. It is interesting to
note that the $0^- 0$ state shows a better convergence and stability
with respect to the $N_{\rm max}$ change as well as a weaker dependence
on $\hbar\Omega$ than the first excited $0^+ 0$ state. This observation
is confirmed also in the point-nucleon rms radius calculation
as can be seen in Fig. \ref{figr93rtpm}. In the experiment,
the $0^- 0$ excitation energy, 21.01 MeV, is higher than the excitation
energy of the first $0^+ 0$ state, 20.21 MeV. Though in our calculations
their positions are reversed, it is visible from Fig. \ref{figr93etpm}
that the extrapolation to larger $N_{\rm max}$ leads to correct
ordering of the two states. A possible interpretation of this observation is
that the excited $0^+ 0$ state is associated with a radial excitation
and, thus, it is more sensitive to the HO basis used in our calculations.

The energy and radius results, 
obtained using the Argonne V8' NN potential,
are presented in Figs. \ref{figv8pe} and \ref{figv8pr}, respectively.
The three-body effective interaction was used in calculating 
these results, for three different HO energies, 
$\hbar\Omega=16, 19$ and 22 MeV. The dotted line 
represents the value -25.92 MeV obtained for the ground state,
using the GFMC \cite{Pipcom}. Similarly, as in the calculations
with the Reid93 NN potential, we get the best convergence for the ground
state for the highest value of $\hbar\Omega$, while for the excited
state the best results are obtained for the lowest $\hbar\Omega$.
The same discussion, given earlier, for the excited-state convergence 
using the Reid93 NN potential, is also valid for the calculations using
the V8' NN potential.
The energy convergence is very slow and there is no sign of convergence 
of the
point-nucleon rms radius of the excited $0^+ 0$ state. 
A significant dependence on $\hbar\Omega$  
prevails for all the model-spaces studied.
On the other hand, the ground-state energy shows 
good convergence and approaches
the GFMC result, in particular for the $\hbar\Omega=22$ MeV calculation.
The ground-state point-nucleon rms radius is almost 
$\hbar\Omega$ independent
and converged. It agrees with the GFMC value of 1.485 fm.   

We note that results on the first excited $0^+ 0$ state obtained
using the resonating group method were reported in Ref. \cite{HH97}.
The Bonn potential employed in that work gives very similar results
for the ground state as those obtained using the Argonne V8'.
It is, therefore, reasonable to make a comparison for the excited
state results. The first excited $0^+ 0$ state energy 
reported in Ref. \cite{HH97} was -6.42 MeV, 
which is about 10\% below our result of $N_{\rm max}=14$
and $\hbar\Omega=16$ MeV calculation. The reported rms radius, 3.02 fm,
is slightly above our calculation.  

In order to further compare the convergence and the $\Omega$ dependence
of the results obtained with two- and three-body effective interactions,
we present a similar calculation as that of Fig. \ref{figr93et}, 
obtained 
using the Argonne V8' NN potential and a larger $\Omega$ difference,
in Fig. \ref{figv8p23}.
The full lines correspond to the three-body effective interaction
calculations, also shown in Fig. \ref{figv8pe}, 
while the dashed lines connect the two-body effective
interaction results. Two HO energies of $\hbar\Omega=$16 and 22 MeV were
used. The dotted line represents the GFMC result. Again we observe
a better stability of the three-body effective-interaction results
with respect to the model-space size changes, a smaller $\Omega$ dependence,
and a faster convergence, in paticular for the ground-state.
In Table \ref{tab0} we show the absolute value of the ground-state 
energy differences obtained in the calculations with HO energies of 
$\hbar\Omega=16$ MeV and $\hbar\Omega=22$ MeV for both
the two-body and the three-body effective-interaction calculations 
in different model spaces. We can see that the differences obtained
with the three-body effective interaction are almost two times smaller
in model spaces with $N_{\rm max}=6-10$. The differences decrease
with the enlargement of the model space for $N_{\rm max}\geq 8$.
We note that by construction the present two- and three-body 
effective-interaction calculations would become identical 
for $N_{\rm max}=28$.

In Table \ref{tab1} we present a summary of our results
obtained in the largest model spaces used in the present study,
e.g., $N_{\rm max}=14$ for the positive-parity states and
$N_{\rm max}=13$ for the negative-parity states. The 
positive-parity state results were obtained using the three-body
effective interaction. For the Argonne V8' NN potential calculations
we also include the GFMC ground-state results \cite{Pipcom} for
a comparison. We note that the Faddeev-Yakubovski equation solution
gives -25.03 MeV \cite{GH93} for the Nijmegen NN potential \cite{NRS78}, 
which gives comparable results to the Reid93 NN potential 
for the three-nucleon problem. The experimental binding energy
of $^4$He is 28.296 MeV. The discrepancy between the experimental
and calculated values are usually attributed to the real three-nucleon 
forces that were not taken into account either in our calculation
or in the other calculations, which we discussed.  
We note that the difference in the binding energies obtained using
the V8' and the Reid93 NN potentials is mainly due to the Coulomb
interaction included in an isospin-invariant manner only in the
calculations with the Reid93 NN potential.

\subsection{Charge form factors}

A sensitive test of the wave-functions obtained in our
calculations is the evaluation of charge form factors. Using the formalism
of Ref. \cite{MSD94}, we calculated the charge EM and strangeness
form factors in the impulse approximation. The one-body contribution
to the charge operator is given by Eq. (15) in Ref. \cite{MSD94}, e.g.,
\begin{equation}\label{chargeop}
\hat{M}_{00}^{(a)}(q)^{[1]}=\frac{1}{2\sqrt{\pi}}\sum_{k=1}^A
\left\{\frac{G_E^{(a)}(\tau)}{\sqrt{1+\tau}}j_0(q r_k)+
\left[ G_E^{(a)}(\tau) - 2 G_M^{(a)}(\tau) \right] 
2\tau \frac{j_1(q r_k)}{q r_k} {\bf \sigma}_k \cdot {\bf L}_k
\right\} \; ,
\end{equation}
where $\tau=\frac{q^2}{4 m_N ^2}$, ${\bf L}_k$ is the $k$-th nucleon
orbital momentum, $G_E^{(a)}(\tau)$ and $G_M^{(a)}(\tau)$ 
are the one-body electric and magnetic form factors, respectively. 
The superscript $(a)$ refers to $(T=0)$ for isoscalar EM form factor 
or to $(s)$ for the strangeness form factor. We use 
the parametrization of the one-body form factors as 
discussed in Ref. \cite{MSD94}
\begin{mathletters}\label{obform}
\begin{eqnarray}
G_E^{(p)}(\tau) &=& G_V^D(\tau) \; , \\
G_M^{(p)}(\tau) &=& \mu_p G_V^D(\tau) \; , \\
G_E^{(n)}(\tau) &=& -\mu_n \tau G_V^D(\tau)\xi_n(\tau) \; , \\
G_M^{(n)}(\tau) &=& \mu_n G_V^D(\tau) \; , \\
G_E^{(s)}(\tau) &=& \rho_s \tau G_V^D(\tau)\xi_s(\tau) \; , \\
G_M^{(s)}(\tau) &=& \mu_s G_V^D(\tau) \; , 
\end{eqnarray}
\end{mathletters} 
with
\begin{mathletters}\label{obformpar}
\begin{eqnarray}
G_V^D(\tau) &=& (1+\lambda^D_V \tau)^{-2} \; , \\
\xi_n &=& (1+\lambda_n \tau)^{-1} \; , \\
\xi_s &=& (1+\lambda^{(s)}_E \tau)^{-1} \; . 
\end{eqnarray}
\end{mathletters} 
The isoscalar EM form factor is given by 
$G_{E,M}^{(T=0)}=\frac{1}{2}[G_{E,M}^{(p)}+G_{E,M}^{(n)}]$,
and for the parameters appearing in Eqs. (\ref{obform},\ref{obformpar}),
one has numerically $\mu_p=2.79, \mu_n=-1.91, 
\lambda_V^D=4.97$, and
$\lambda_n = 5.6$. Following Ref. \cite{MSD94}, we also set
the strangeness radius $\rho_s =-2.0$ and $\lambda^{(s)}_E = \lambda_n$.
Limits on these parameters are to be determined in the experiments
at the Thomas Jefferson Accelerator Facility (TJNAF). 
The first strangeness magnetic-moment measurement
was reported recently \cite{M97} and an experimental value 
$\mu_s=+0.23$,
obtained with a large error. We use this value in our calculations.

Our charge form factor calculations are presented in Figs. \ref{figEM22}
-\ref{figratEMS22}. The charge form factors given in the figures
were calculated using the one-body operator (\ref{chargeop})
as $F_C^{(a)}(q)=2\sqrt{\pi}\langle f, 0^+ 0|\hat{M}_{00}^{(a)}(q)^{[1]}
|i,0^+ 0\rangle$.
We show only results obtained with the Argonne V8' NN potential;
the Reid93 NN potential gives almost identical results for
the charge form factors, when the same HO energy
$\hbar\Omega$ is employed. Our calculated elastic EM charge form factor 
is given in Fig. \ref{figEM22} together with the inelastic EM charge form 
factor corresponding to the transition to the first excited 
$0^+ 0$ state.
These results were obtained using the HO energy
$\hbar\Omega=22$ MeV and the three-body effective interaction
in the $N_{\rm max}=14$ model space. In this calculation
we obtained the best description of the ground state. 
The calculation of the elastic charge form factor in the impulse
approximation can be directly compared to that presented 
in Fig. 2 of Ref. \cite{MSD94}, performed using Variational
Monte Carlo (VMC) wave-functions and the Argonne V14 NN potential.
There, the minimum was obtained at
$q\approx 3.55$ fm$^{-1}$, while the experimental minimum is 
at $q\approx 3.2$ fm$^{-1}$. The difference can be explained
with the help of meson-exchange-current contributions. 
The elastic charge EM form factor obtained in our calculation 
compares well with that obtained by the VMC wave-functions.
It is shifted further
to higher $q$, namely, we get the minimum at $q\approx 3.75$ fm$^{-1}$. 
We note that a second minimum appears in our calculated
elastic charge form factor at $q\approx 7.25$ fm$^{-1}$.
The second minimum at a similar position was found in the
VMC calculations presented in Ref. \cite{SPR90}.   
To examine the form factor dependence on $\hbar\Omega$, 
we repeated these calculations for different choices 
of the HO energy. In Fig. \ref{figEM19} 
we show the result obtained with $\hbar\Omega=19$ MeV.
All other characteristics are the same as in the calculation 
of Fig. \ref{figEM22}. The minimum here is shifted further
to higher $q$, we have it at $q\approx 3.85$ fm$^{-1}$. 
The difference between the two results is rather small
but still it 
shows that our calculation is not completely converged
and, in particular, the description of the high 
transferred-momentum part of the form factors requires 
the use of even larger model spaces than we employed.   
We note that the inelastic form factor has a stronger
$\Omega$ dependence than the elastic form factor.
As discussed in the previous subsection, convergence 
of the excited state has not been achieved in our calculations
within the model spaces employed. Therefore, our calculated 
inelastic form factors must be taken with some degree of caution.
Let us remark that, in addition to the transition form factor,
we also computed the form factor of the first excited $0^+ 0$
state. That form factor was also evaluated in Ref. \cite{HH97}
in the resonating group method approach using the Bonn potential.
Similarly as in that work, our calculated $0^+_2$ form factor
is almost an order of magnitude smaller than the ground-state
form factor for a wide range of $q$. 
The first minimum is shifted in our calculation
to larger $q$, more or less to the position of the transition 
form factor minimum, and the second minimum is shifted 
to smaller $q$ compared to our ground-state form factor.
 
Our calculated elastic strangeness form factor
together with the inelastic EM charge form 
factor corresponding to the transition to 
the first excited $0^+ 0$ state are shown in Fig. \ref{figS22}.
These results were obtained using the same wave-functions as
those used for calculations presented in Fig. \ref{figEM22},
namely we had $\hbar\Omega=22$ MeV, the model-space size
characterized by $N_{\rm max}=14$ and the three-body effective 
interaction was employed. 
The elastic form factor can be compared with the impulse 
approximation VMC result of Fig. 3 in Ref. \cite{MSD94}.
Similarly, as for the EM elastic form factor, our calculation 
compares well with VMC result.
We note, however, the different value of strangeness magnetic moment
used in Ref. \cite{MSD94} ($\mu_s=-0.2$).

Finally, in Fig. \ref{figratEMS22} we present the ratio of the
EM and strangeness form factors from Figs. \ref{figEM22} and \ref{figS22}.
The ratio of the elastic charge form factors is particularly interesting,
as it can be experimentaly obtained from the measurement of the
parity-violating left-right asymmetry for scattering of polarized
electrons from a $^4$He target. Experiments of this type are now under
preparation at TJNAF.

\section{Conclusions}
\label{sec4}

In the present study we used equations for Faddeev amplitudes,
antisymmetrized for three nucleons, to solve the shell-model
problem for the four-nucleon system. We performed calculations
in larger model spaces, up to an HO excitation of $14\hbar\Omega$ 
above the unperturbed ground-state, than in any other
shell-model study so far. The main motivation for the present work
was to test the shell-model approach and the effective 
interactions that we want to apply to more complex systems,
e.g., $p$-shell nuclei, in particular. The effective interactions that
we employed were derived from realistic NN potentials, i.e.,
the Reid93 and the Argonne V8'. In addition to the two-body effective
interactions, we also computed the three-body effective interactions
and demonstrated that their use significantly improves 
the convergence of the results. 

Our calculations depend on the model-space
size and on the HO frequency $\Omega$.
The effective interactions were constructed in such a way that
in the large model-space limit the effective interactions approach
the bare NN interaction. Thus our results should converge to the exact
solutions. The dependence on the model-space size and $\Omega$
was investigated. We found quite different behavior of the ground
state and the first excited $0^+ 0$ state. Our ground-state energy
and point-nucleon radius results begin to converge and are close 
to or in agreement with those obtained by the GFMC method.
For the first excited $0^+ 0$ state, our results, the point-nucleon radius,
in particular, show large model-space and $\Omega$ dependence.
This implies that significantly larger model spaces would still
be needed in order to obtain the exact solutions. The nature
of the $0_2^+$ state is discussed in the literature \cite{CVS88,CH97}.
The Coulomb interaction plays an important role in the description
of this state. In the present calculations we did not include 
the isospin breaking. Our formalism is quite general, however, 
and allows the use of interactions that break the isospin symmetry.
On the other hand, the calculated properties of the $0^-_1$ 
state show better convergence behavior. In the model spaces studied,
we obtained lower excitation energy of the $0^-_1$ state than
of the $0^+_2$ state, contrary to experiment. The 
extrapolation of the model-space dependence of these two energies
to larger model spaces shows, however, that the correct ordering 
of the states will be obtained. Apparently,
the $0^+_2$ state is associated with a radial excitation
and, thus, it is more sensitive to the HO basis used in the expansion.

A sensitive test of our calculated wave functions is the computation
of the charge EM and strangeness form factors. Our impulse-approximation 
results show little dependence on the NN potential and our best 
results are
close to the corresponding form factors obtained using the VMC
wave functions and the Argonne V14 NN potential. 
In particular, we observe both the first and the second minima 
in the elastic charge
form factor in positions close to those obtained using the VMC 
calculations.
In addition
to the elastic charge form factors, we also evaluated the form factors
for the transition to the $0^+_2$ state in the impulse approximation.

In general, the energy scales of the bound $0s$ nucleons are significantly
different from the scattering energies of the resonances. This difference
can only be accounted for with a large $N_{\rm max}$ in our approach.
Consequently, the results for the excited states and the transition
form factors obtained within the limited model spaces of
the present work should be taken with some caution.
In the future we would like to apply the formalism, discussed 
in the present paper, 
to a more extensive study of the negative-parity states of $^4$He.
In particular, it is desirable to use still larger model spaces
to investigate the excited-state convergence properties.

The most important result of the present work is, however, 
the successful use of the three-body effective interaction. 
This three-body effective interaction can be computed
for more complex nuclei as well and, in principle, used,
after a transformation to an appropriate three-nucleon basis, 
in standard shell-model calculations. A more practical
approach, however, is to make use of the three-body effective
interaction knowledge for the renormalization of the two-body
effective interaction. Work in this direction is under way.
In addition, the present formalism may be used to compute
the four-body effective interaction for nuclei with $A>4$. 
We plan to extend the shell-model Faddeev-like approach
that we have successfully applied to three- and four-nucleon systems
to systems with more than four nucleons, using also a formalism
of equations for components with lower degree of antisymmetry
than the full wave-function developed in Ref. \cite{GPK}.

\acknowledgements{
We thank Gintautas P. Kamuntavi\v{c}ius for useful comments 
concerning the symmetries present in Eqs. (\ref{t13t23}) and (\ref{t4}).
This work was supported in part by the NSF grant No. PHY96-05192.
P.N. also acknowledges partial support from the grant of the 
Grant Agency of the Czech Republic 202/96/1562. 
}

\begin{figure}
\caption{The dependence of the ground-state and the 
first-excited $0^+ 0$ state energies, in MeV, 
on the maximal number of HO excitations allowed
in the model space. The two-body effective interaction 
utilized was derived
from the Reid93 NN potential. 
Results for $\hbar\Omega=14, 17, 19$ and 22 MeV are presented.
}
\label{figr93e}
\end{figure}

\begin{figure}
\caption{The dependence of the point-nucleon 
rms radius of the ground state and 
the first-excited $0^+ 0$ state, in fm, 
on the maximal number of HO excitations allowed
in the model space. The two-body effective interaction 
utilized was derived
from the Reid93 NN potential. 
Results for $\hbar\Omega=14, 17, 19$ and 22 MeV are presented.
}
\label{figr93r}
\end{figure}

\begin{figure}
\caption{The dependence of the 
ground-state and the first-excited $0^+ 0$ state energies, in MeV, 
on the maximal number of HO excitations allowed
in the model space. Results obtained using the two-body (dashed line) 
and three-body (full line) effective interaction derived 
from the Reid93 NN potential are compared. 
Harmonic-oscillator energies of $\hbar\Omega=17$ and 19 MeV were used.
}
\label{figr93et}
\end{figure}

\begin{figure}
\caption{The dependence of the 
point-nucleon rms radius of the ground state and 
the first-excited $0^+ 0$ state, in fm, 
on the maximal number of HO excitations allowed
in the model space. Results obtained using the two-body (dashed line) 
and three-body (full line) effective interaction derived 
from the Reid93 NN potential are compared. 
Harmonic-oscillator energies of $\hbar\Omega=17$ and 19 MeV were used.
}
\label{figr93rt}
\end{figure}

\begin{figure}
\caption{The dependence of the 
ground state, the first-excited $0^+ 0$ state 
and the first-excited $0^- 0$ state energies, in MeV, 
on the maximal number of HO excitations allowed
in the model space. 
For the positive-parity states the results were obtained 
using the three-body effective interaction derived 
from the Reid93 NN potential. Energies of the $0^- 0$ state
were calculated using a two-body effective interaction derived
from the Reid93 NN potential. 
Harmonic-ocillator energies of $\hbar\Omega=17$ and 19 MeV were used.
}
\label{figr93etpm}
\end{figure}

\begin{figure}
\caption{The dependence of the 
point-nucleon rms radius of the 
ground-state, the first-excited $0^+ 0$ state 
and the first-excited $0^- 0$ state, in fm, 
on the maximal number of HO excitations allowed
in the model space. 
For the positive-parity states the results were obtained 
using the three-body effective interaction derived 
from the Reid93 NN potential. Energies of the $0^- 0$ state
were calculated using a two-body effective interaction derived
from the Reid93 NN potential. 
Harmonic-oscillator energies of $\hbar\Omega=17$ and 19 MeV were used.
}
\label{figr93rtpm}
\end{figure}

\begin{figure}
\caption{The dependence of the 
ground state and the first-excited $0^+ 0$ state energies, in MeV, 
on the maximal number of HO excitations allowed
in the model space. The three-body effective interaction derived
from the Argonne V8' NN potential was used. 
Results for $\hbar\Omega=16, 19$ and 22 MeV are presented.
The dotted line represents the result -25.92 MeV of the GFMC
calculation \protect{\cite{Pipcom}}.
}
\label{figv8pe}
\end{figure}

\begin{figure}
\caption{The dependence of the 
Point-nucleon rms radius of the ground state and 
the first-excited $0^+ 0$ state, in fm, 
on the maximal number of HO excitations allowed
in the model space. The three-body effective interaction derived
from the Argonne V8' NN potential was used. 
Results for $\hbar\Omega=16, 19$ and 22 MeV are presented.
}
\label{figv8pr}
\end{figure}

\begin{figure}
\caption{The dependence of the 
ground-state and the first-excited $0^+ 0$ state energies, in MeV, 
on the maximal number of HO excitations allowed
in the model space. Results obtained using the two-body (dashed line) 
and three-body (full line) effective interaction derived 
from the Argonne V8' NN potential are compared. 
Harmonic-oscillator energies of $\hbar\Omega=16$ and 22 MeV were used.
The dotted line represents the result -25.92 MeV of the GFMC
calculation \protect{\cite{Pipcom}}.
}
\label{figv8p23}
\end{figure}

\begin{figure}
\caption{The elastic EM charge form factor (full line) and the EM charge
form factor corresponding to the transition to  
the first-excited $0^+ 0$ state (dotted line) calculated 
in the impulse approximation using the three-body effective 
interaction derived from the Argonne V8' NN 
potential in the $N_{\rm max}=14$ model space and $\hbar\Omega=22$ MeV.
}
\label{figEM22}
\end{figure}

\begin{figure}
\caption{The elastic EM charge form factor (full line) and the EM charge
form factor corresponding to the transition to  
the first-excited $0^+ 0$ state (dotted line) calculated 
in the impulse approximation using the three-body effective 
interaction derived from the Argonne V8' NN
potential in the $N_{\rm max}=14$ model space and $\hbar\Omega=19$ MeV.
}
\label{figEM19}
\end{figure}

\begin{figure}
\caption{The elastic strangeness charge form factor (full line) 
and the strangeness charge form factor corresponding to the transition to  
the first-excited $0^+ 0$ state (dotted line) calculated 
in the impulse approximation using the three-body effective 
interaction derived from the Argonne V8' NN
potential in the $N_{\rm max}=14$ model space and $\hbar\Omega=22$ MeV.
Values of the strangeness radius $\rho_s=-2.0$ and 
the strangeness magnetic moment
$\mu_s=0.23$ were employed.
}
\label{figS22}
\end{figure}

\begin{figure}
\caption{The ratio of the elastic EM and strangeness charge form 
factors (full line) 
and the EM and strangeness charge form factor 
corresponding to the transition to  
the first-excited $0^+ 0$ state (dotted line) calculated 
in the impulse approximation using the three-body effective 
interaction derived from the Argonne V8' NN
potential in the $N_{\rm max}=14$ model space and $\hbar\Omega=22$ MeV.
Values of the strangeness radius $\rho_s=-2.0$ and 
the strangeness magnetic moment
$\mu_s=0.23$ were employed.
}
\label{figratEMS22}
\end{figure}

\begin{table}
\begin{tabular}{cccccc}
$N_{\rm max}$ & 6 & 8 & 10 & 12 & 14 \\
\hline
$|\Delta E_{\rm 2eff}|$ & 1.311 & 1.466 & 1.265 & 1.037 & 0.834 \\
$|\Delta E_{\rm 3eff}|$ & 0.778 & 0.782 & 0.676 & 0.601 & 0.550
\end{tabular}
\caption{Absolute value of the ground-state energy differences obtained
in the calculations with HO energies of 
$\hbar\Omega=16$ MeV and $\hbar\Omega=22$ MeV
with the two-body (second row) and the 
three-body (third row) effective interactions in different model spaces.
The effective interactions were derived from the Argonne V8' NN potential.
The corresponding energy dependence is shown in 
Fig.~\protect{\ref{figv8p23}}.   
}
\label{tab0}
\end{table}

\begin{table}
\begin{tabular}{cccccc}
Argonne V8' NN potential & &&&& \\
\hline
State& Variable& $\hbar\Omega=16$ MeV &$\hbar\Omega=19$ MeV &
$\hbar\Omega=22$ MeV & GFMC \\
$0^+_1$ & $E$ [MeV] & -26.62 & -26.30 & -26.07 & -25.92(8) \\
        & $\sqrt{\langle r^2 \rangle}$ [fm] 
                    & 1.481 & 1.485 & 1.485 & 1.485(10) \\
$0^+_2$ & $E$ [MeV] & -5.77 & -4.89 & -3.93 &  \\
        & $E_x$ [MeV] & 20.86 & 21.42 & 22.14 & \\
        & $\sqrt{\langle r^2 \rangle}$ [fm] 
                    & 2.906 & 2.777 & 2.658 & \\
$0^-_1$ & $E$ [MeV] & -6.70 & -6.17 & -5.59 &\\
        & $E_x$ [MeV] & 19.93 & 20.14 & 20.48 & \\
        & $\sqrt{\langle r^2 \rangle}$ [fm] 
                   & 2.349  & 2.263  & 2.186 &  \\ 
\hline
\hline
Reid93 NN potential &&&&& \\
\hline
State& Variable& $\hbar\Omega=17$ MeV &$\hbar\Omega=19$ MeV &&  \\
$0^+_1$ & $E$ [MeV] & -25.69 & -25.47 &&\\
        & $\sqrt{\langle r^2 \rangle}$ [fm] 
                    & 1.487 & 1.489 &&\\
$0^+_2$ & $E$ [MeV] & -5.00 & -4.39 &&\\
        & $E_x$ [MeV] & 20.69 & 21.08 &&\\
        & $\sqrt{\langle r^2 \rangle}$ [fm] 
                    & 2.873 & 2.787 &&\\
$0^-_1$ & $E$ [MeV] & -5.91 & -5.54  &&\\
        & $E_x$ [MeV] & 19.78 & 19.93  && \\
        & $\sqrt{\langle r^2 \rangle}$ [fm] 
                    & 2.339  &  2.281   && 
\end{tabular}
\caption{Results for the ground-state and excited state 
energies and point-nucleon 
rms radii, as well as the excitation energies ($E_x$),  
obtained in the largest model spaces used in the present study,
$N_{\rm max}=14$, (13) for the positive- (negative-)parity states,
respectively, are presented. 
All the states have isospin $T=0$.
The positive-parity-state calculations were performed 
using the three-body effective interaction. 
Results for different HO energies are given in
separate columns.
The GFMC ground-state
results \protect\cite{Pipcom} are shown for comparison.}
\label{tab1}
\end{table}


\begin{references}

\bibitem {Fad60} L.D. Faddeev, Zh. Eksp. Teor. Fiz. {\bf 39},
                 1459 (1960) [Sov. Phys.-JETP {\bf 12}, 1014 (1961)].

\bibitem {PFGA80} G.L. Payne, J.L. Friar, B.F. Gibson, and I.R. Afnan, 
         Phys. Rev. {\bf C 22}, 823 (1980).

\bibitem {PFG80} G.L. Payne, J.L. Friar, and B.F. Gibson,
         Phys. Rev. {\bf C 22}, 832 (1980).

\bibitem {CPFG85} C.R. Chen, G.L. Payne, J.L. Friar, and B.F. Gibson,
         Phys. Rev. {\bf C 31}, 2266 (1985).

\bibitem {FPSS93} J.L. Friar, G.L. Payne, V.G.J. Stoks,
             and J.J. de Swart,
         Phys. Lett. {\bf B 311}, 4 (1993).

\bibitem {NHKG97} A. Nogga, D. H\"uber, H. Kamada, and W. Gl\"ockle,
                  Phys. Lett. B {\bf 409}, 19 (1997).   

\bibitem {Ya67} O. A. Yakubovsky, Sov. J. Nucl. Phys. {\bf 5},
	        937 (1967).

\bibitem {GH93} W. Gl\"ockle and H. Kamada, Phys. Rev. Lett.
                {\bf 71}, 971 (1993).

\bibitem{GFMC} B. S. Pudliner, V. R. Pandharipande, J. Carlson, 
               S. C. Pieper and R. B. Wiringa, 
               Phys. Rev. {\bf C 56} 1720, (1997);
               R. B. Wiringa, Nucl. Phys. {\bf A 631}, 70c (1998).

\bibitem {VKR95} M. Viviani, A. Kievsky, and S. Rosati,
                Few-Body Systems {\bf 18}, 25 (1995).   
         
\bibitem {ZBVHS} D. C. Zheng, J. P. Vary, and B. R. Barrett, Phys. Rev. 
                 {\bf C 50}, 2841 (1994);
		 D. C. Zheng, B. R. Barrett, J. P. Vary, 
                 W. C. Haxton, and C. L. Song, Phys. Rev. 
                 {\bf C 52}, 2488 (1995).

\bibitem {NB96} P. Navr\'atil and B. R. Barrett, 
             Phys. Rev. {\bf C 54}, 2986 (1996);
             Phys. Rev. {\bf C 57}, 3119 (1998).

\bibitem {NB98} P. Navr\'atil and B. R. Barrett, 
             Phys. Rev. {\bf C 57}, 562 (1998).

\bibitem {CVS88} R. Ceuleneer, P. Vandepeutte, and C. Semay,
                 Phys. Rev {\bf C 38}, 2335 (1988). 

\bibitem {CPW84} J. Carlson, V. R. Pandharipande, and R. B. Wiringa,
	       Nucl. Phys. {\bf A424}, 47 (1984).

\bibitem {CH97} A. Cs\`ot\`o and G. M. Hale, Phys. Rev. {\bf C 55},
               2366 (1997).

\bibitem {HH97} H. M. Hofmann and G. M. Hale,
	       Nucl. Phys. {\bf A613}, 69 (1997). 

\bibitem {VRK98} M. Viviani, S. Rosati, and A. Kievsky, 
	        Phys. Rev. Lett. {\bf 81}, 1580 (1998).  

\bibitem {CC98} F. Ciesielski and J. Carbonell, 
	       Phys. Rev. {\bf C 58}, 58 (1998);
	        F. Ciesielski, J. Carbonell, and C. Gignoux,
               Nucl. Phys. {\bf A631}, 653c (1998).     

\bibitem {Tr72} L. Trlifaj, Phys. Rev. {\bf C 5}, 1534 (1972).

\bibitem {HKT72} E. P. Harper, Y. E. Kim, and A. Tubis, Phys. Rev.
                 {\bf C 6}, 126 (1972).

\bibitem {Glo83} W. Gl\"ockle, {\it The Quantum Mechanical Few-Body
                 Problem} (Springer-Verlag, New York, 1983).

\bibitem {RY95} V. A. Rudnev and S. L. Yakovlev, Phys. of At. Nuclei 
	        {\bf 58}, 1662 (1995).

\bibitem {YLV60} A. P. Yutsis, I. B. Levinson, and V. V. Vanagas,
                 Mathematical Apparatus of the Theory of Angular
                 Momentum, (Israel Program for scientific Translations, 
                 Jerusalem 1962). 

\bibitem {LS80} K. Suzuki and S.Y. Lee, Prog. Theor. Phys. {\bf 64}, 
                2091 (1980).

\bibitem {S82SO83} K. Suzuki, Prog. Theor. Phys. {\bf 68},
              246 (1982); 
            K. Suzuki and R. Okamoto, Prog. Theor. Phys. {\bf 70},
              439 (1983).

\bibitem {SKTS} V. G. J. Stoks, R. A. M. Klomp, C. P. F. Terheggen, 
             and J. J. de Swart, Phys. Rev. {\bf C 49} 2950 (1994).

\bibitem {HG83} A. G. M. van Hees and P. W. M. Glaudemans, Z. f. Phys. 
              {\bf A314}, 323 (1983).

\bibitem {VZ94} J. P. Vary and D. C. Zheng, ``The Many-Fermion-Dynamics
            Shell-Model Code'', Iowa State University (1994)
            (unpublished).

\bibitem {NBG98} P. Navr\'atil, B. R. Barrett, and W. Gl\"ockle,
               Phys. Rev. {\bf C 59}, (February 1999). 

\bibitem {VOS} K. Varga and Y. Suzuki, Phys. Rev C {\bf 52},
                 2885 (1995);  
	         K. Varga, Y. Ohbayasi and Y. Suzuki, 
                 Phys. Lett. B {\bf 396}, 1 (1997).

\bibitem {ZBVC} D.C. Zheng, B.R. Barrett, J.P. Vary and
      R.J. McCarthy, Phys. Rev. {\bf C 49}, 1999 (1994).

\bibitem {ZBVM} D.C. Zheng, B.R. Barrett, J.P. Vary and
      H. M\"{u}ther, Phys. Rev. {\bf C 51}, 2471 (1995).

\bibitem {Pipcom} S. C. Pieper, private communication.

\bibitem {NRS78} M. M. Nagels, T. A. Rijken, and J. J. de Swart,
                 Phys. Rev. {\bf D 17}, 768 (1978). 

\bibitem {MSD94} M. J. Musolf, R. Schiavilla, and T. W. Donnelly,
                Phys. Rev. {\bf C 50}, 2173 (1994). 

\bibitem {M97} B. Mueller et al., Phys. Rev. Lett. {\bf 78}, 3824 (1997). 

\bibitem {SPR90} R. Schiavilla, V. R. Pandharipande, and D. O. Riska,
                 Phys. Rev. {\bf C 41}, 309 (1990).

\bibitem {GPK} G. P. Kamuntavi\v{c}ius, Few-Body Systems {\bf 1}, 91
                 (1986); Sov. J. Part. Nuclei, {\bf 20}, 109 (1988).


\end{references}
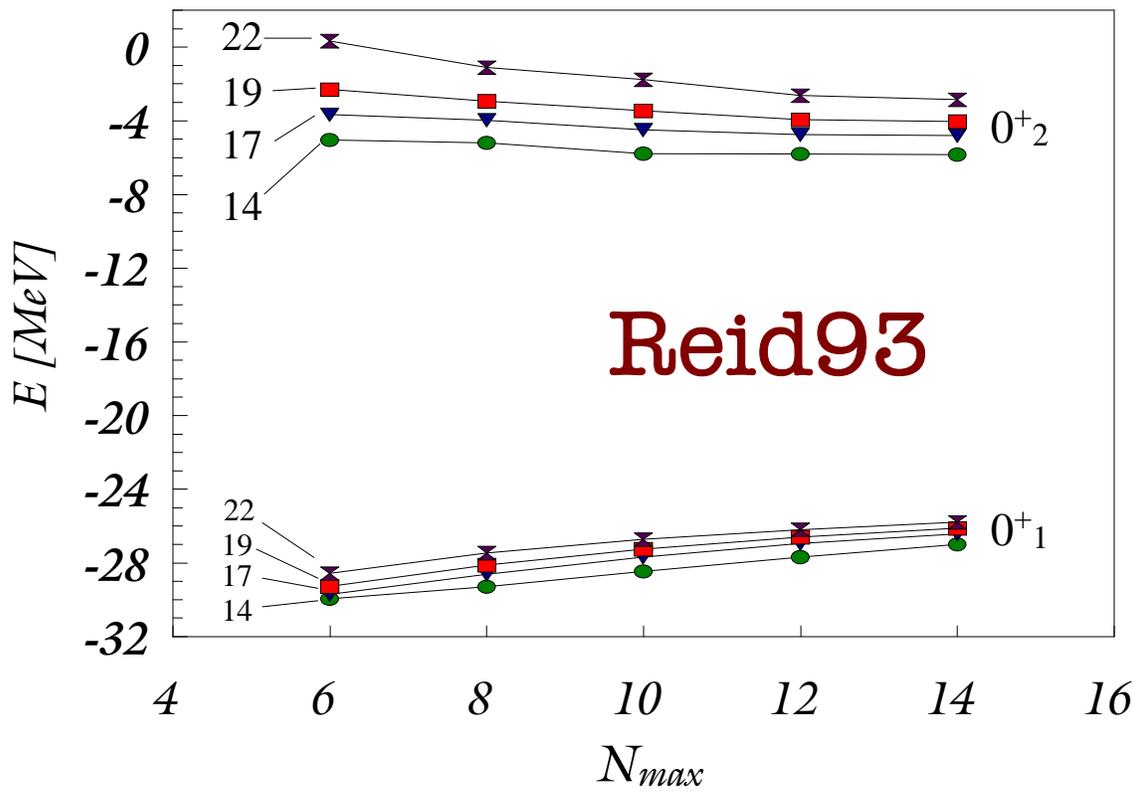
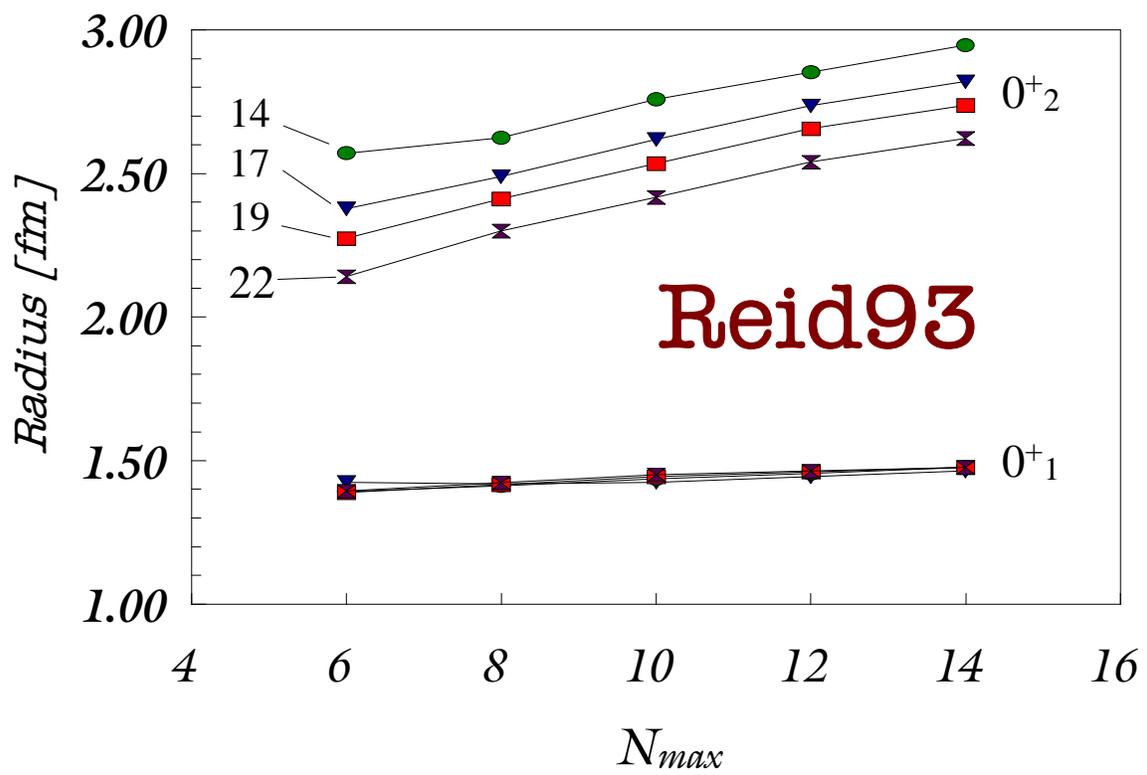
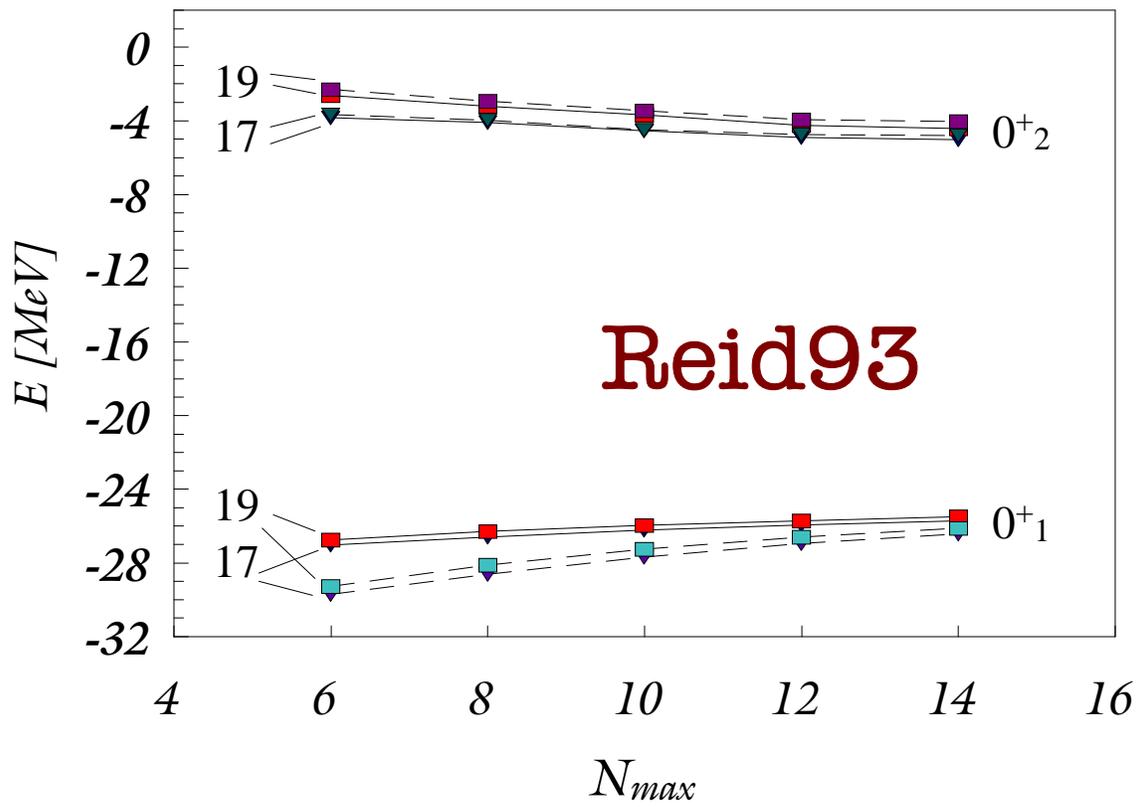
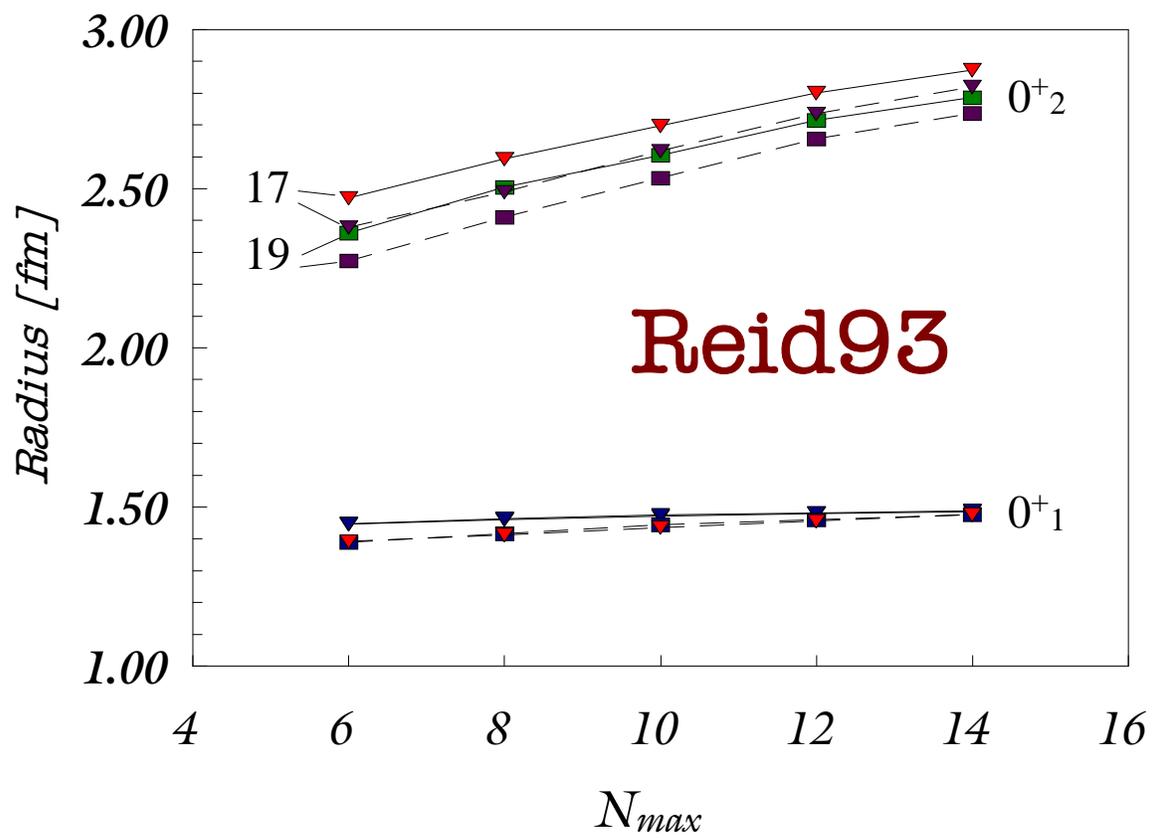
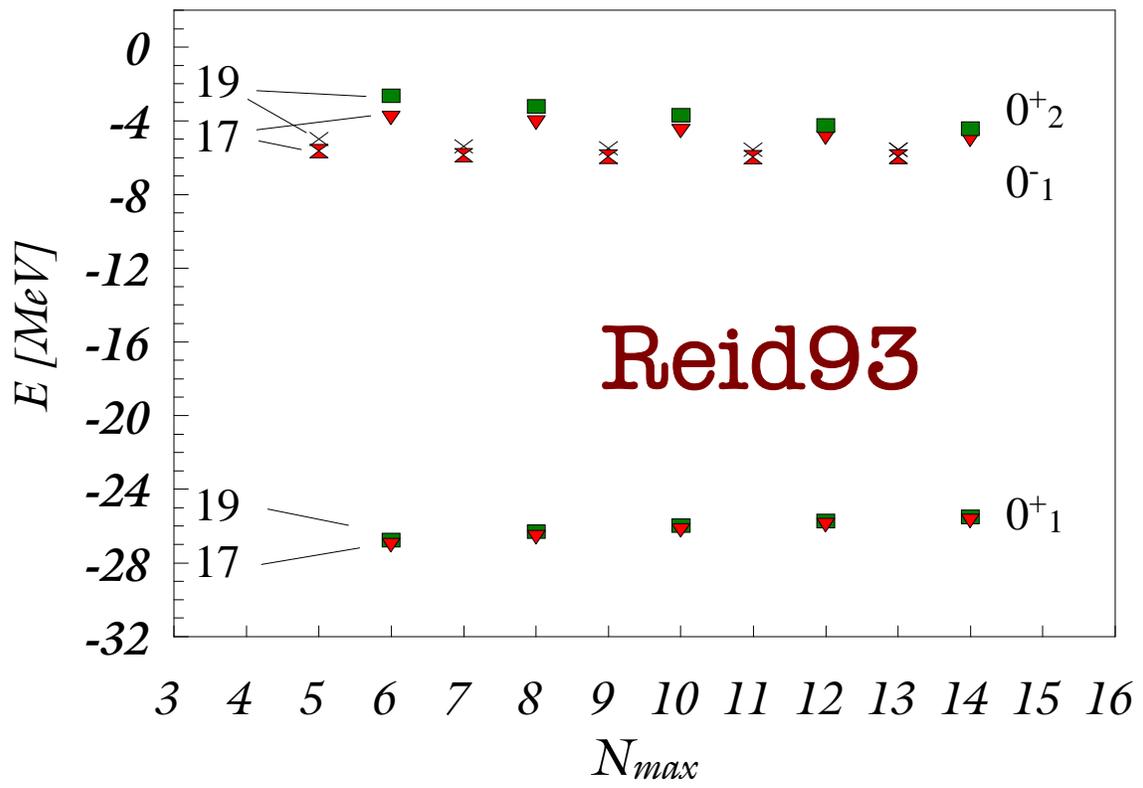
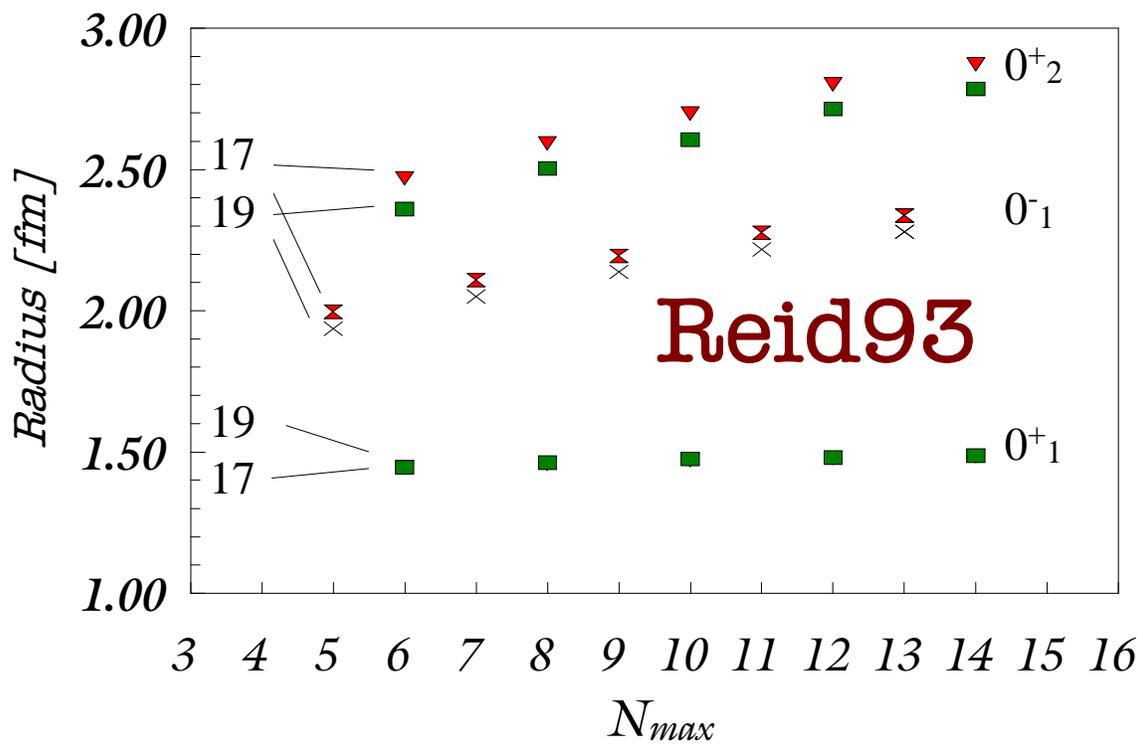
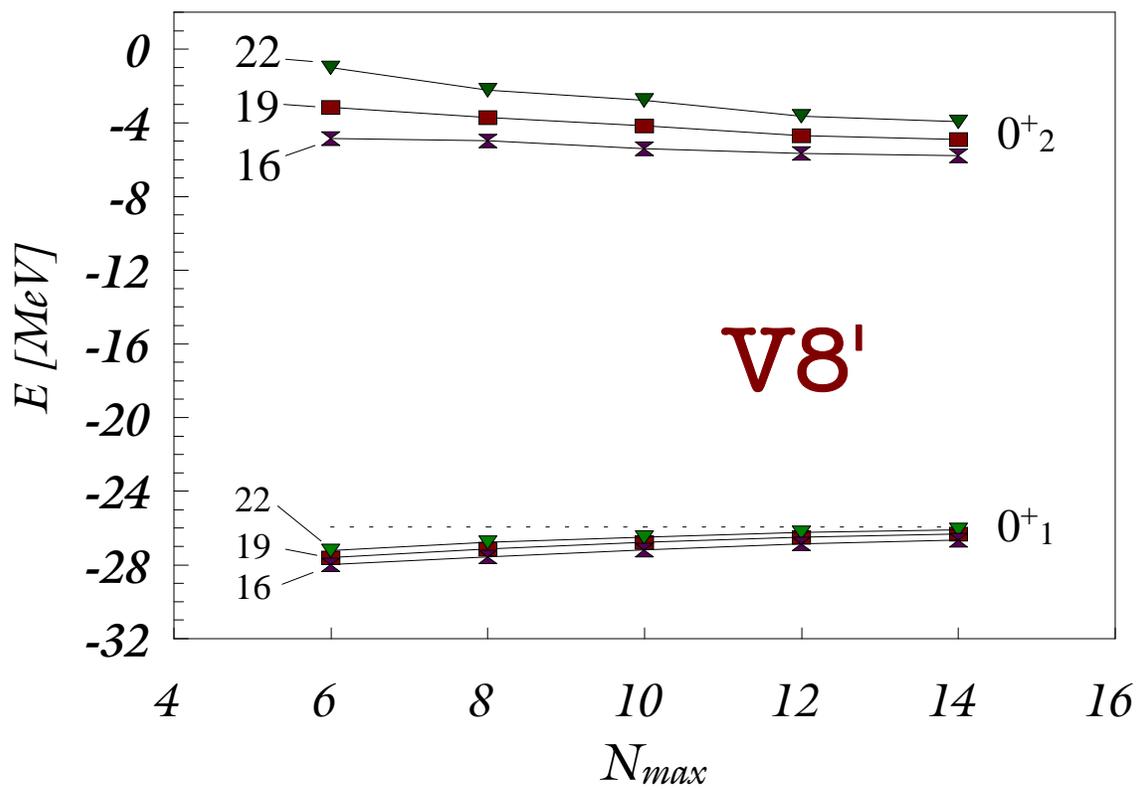
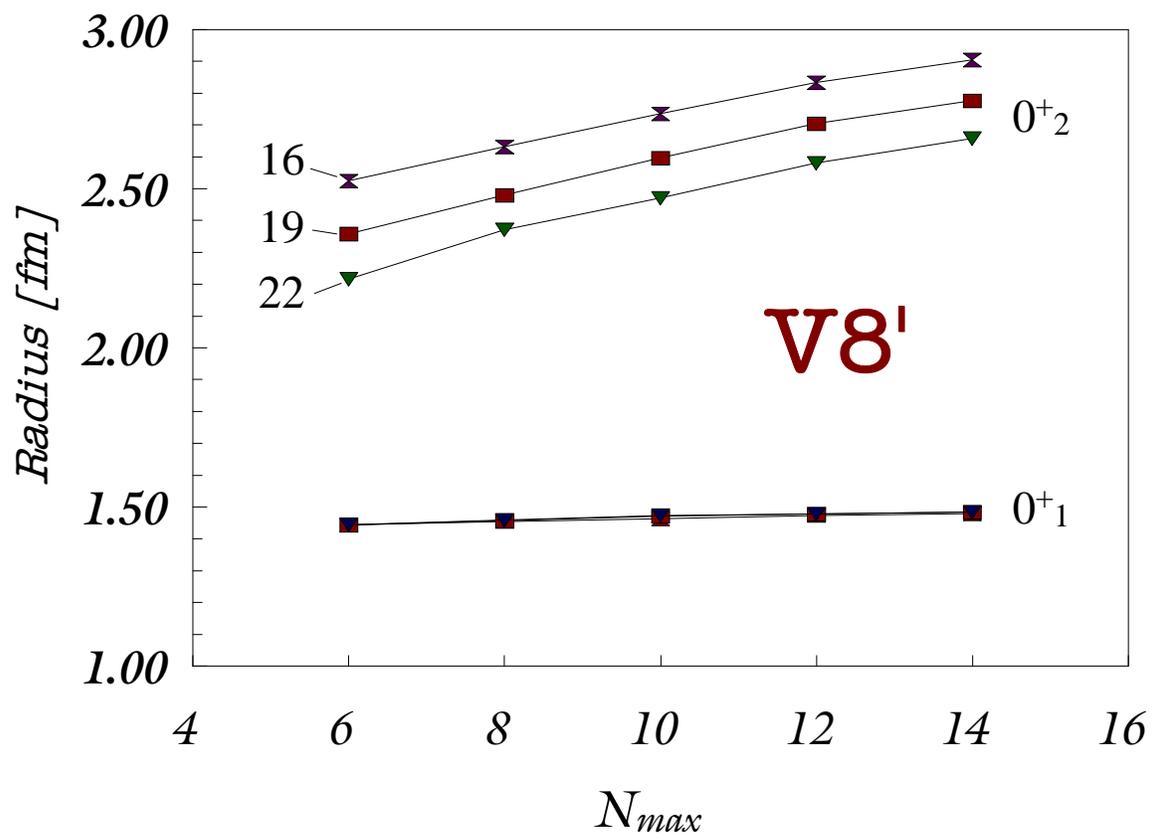
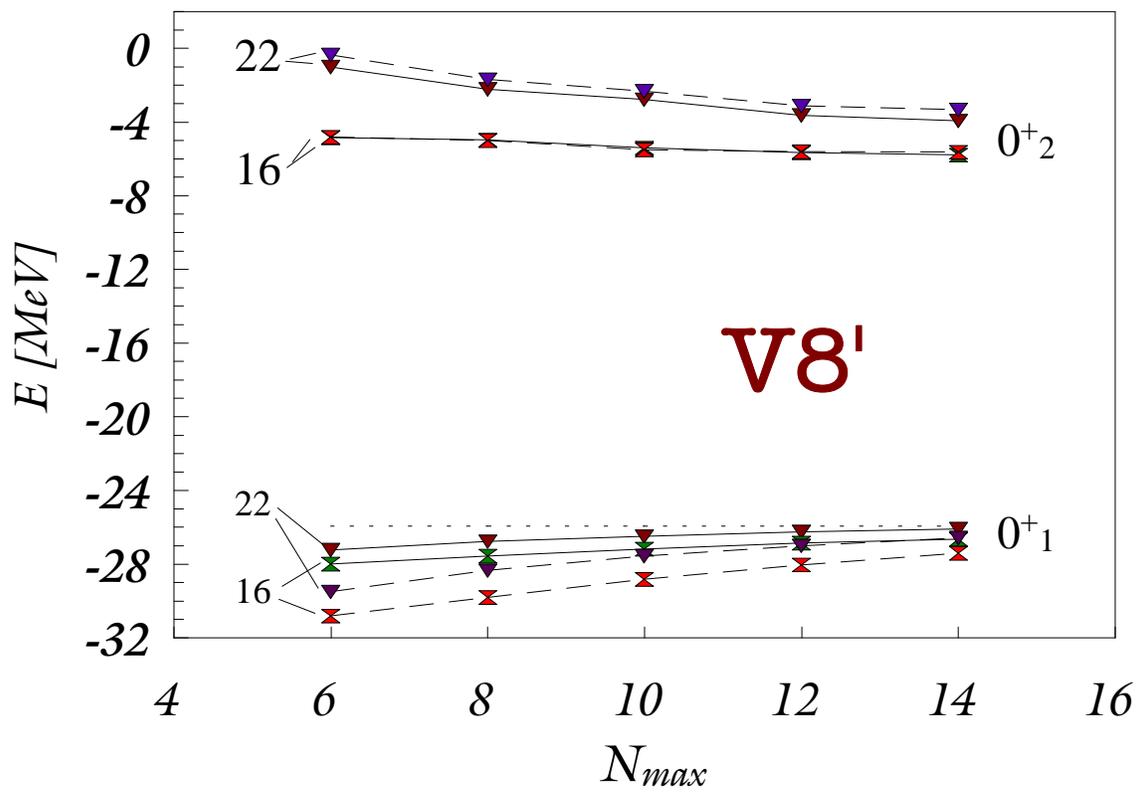
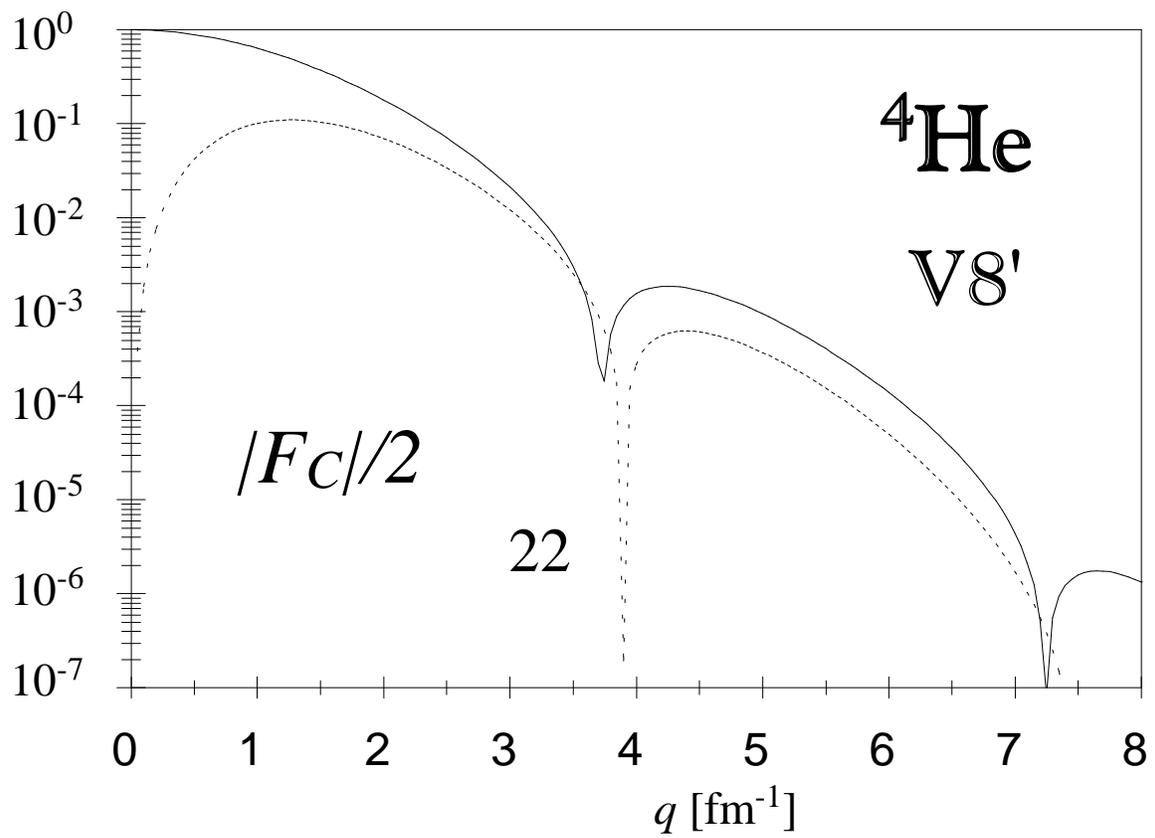
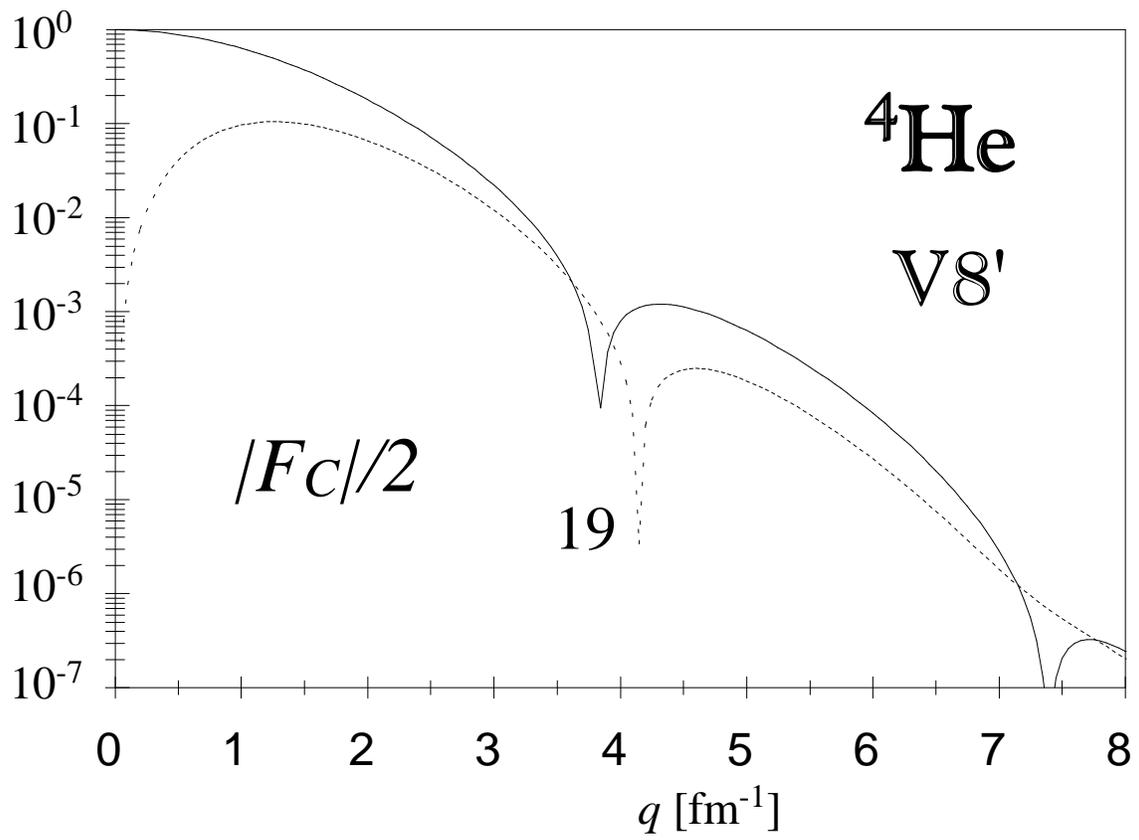
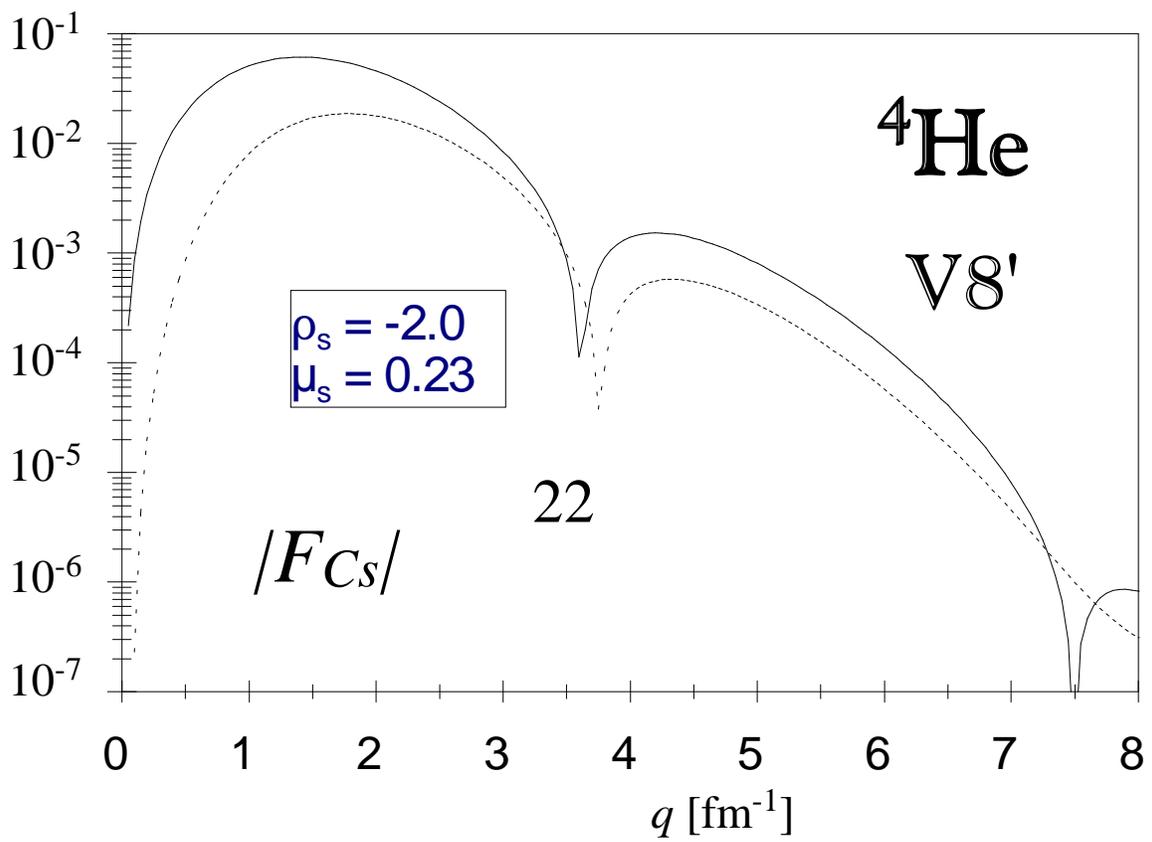
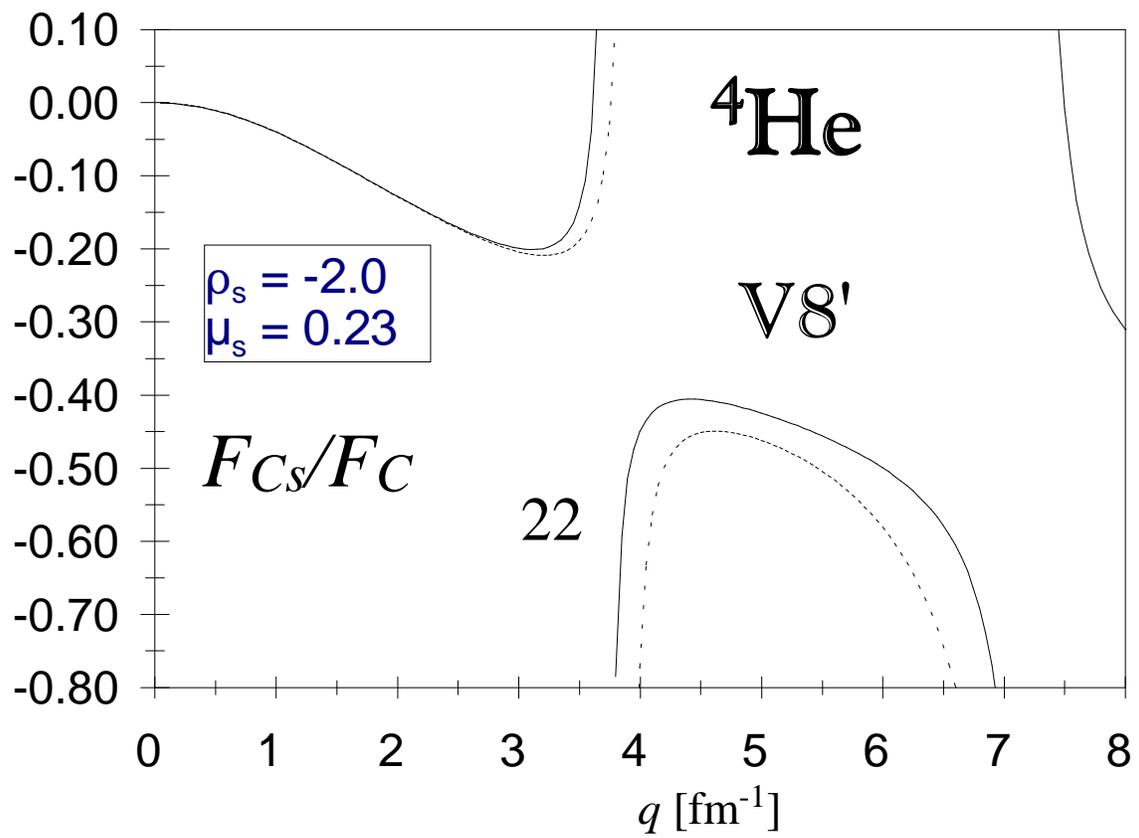
\end{document}